%% file: paper-olga.tex
\documentclass[a4paper,fleqn,usenatbib,useAMS]{mnras}

\usepackage{color}

\usepackage{graphicx}	
\usepackage{amsmath}	
\usepackage{amssymb}	
\usepackage{multicol}        
\usepackage{bm}		
\usepackage{pdflscape}	

\DeclareMathOperator{\sign}{sign}
\defcitealias{yan_lazarian_2011}{YL11}
\definecolor{darkgreen}{RGB}{0,130,0}
\definecolor{darkgreen2}{RGB}{0,80,0}
\definecolor{darkblue}{RGB}{0,0,130}

\usepackage[T1]{fontenc}
\usepackage{ae,aecompl}

\title[CR gyroresonance instability]{Kinetic-MHD simulations of gyroresonance instability driven by CR pressure anisotropy}

\author[O. Lebiga, R. Santos-Lima \& H. Yan]{
	O. Lebiga,$^{1,2,3}$\thanks{Contact e-mail: \href{mailto:olga.lebiga@desy.de}{olga.lebiga@desy.de}}
	R. Santos-Lima,$^{1,2}$\thanks{Contact e-mail: \href{mailto:reinaldo.santos.de.lima@desy.de}{reinaldo.santos.de.lima@desy.de}}
	H. Yan,$^{1,2}$\thanks{Contact e-mail: \href{mailto:huirong.yan@desy.de}{huirong.yan@desy.de}}\\
$^{1}$DESY, Platanenallee 6, 15738 Zeuthen, Germany \\
$^{2}$Institut f\"{u}r Physik und Astronomie, Universit\"{a}t Potsdam, 14476 Potsdam-Golm, Germany \\
$^{3}$Taras Shevchenko National University of Kyiv, 64/13, Volodymyrska Street, Kyiv, Ukraine, 01601 }

\pubyear{2017}

\begin{document}
\label{firstpage}
\pagerange{\pageref{firstpage}--\pageref{lastpage}}
\maketitle

\begin{abstract}
The transport of cosmic rays (CRs) is crucial for 
the understanding of almost all high-energy phenomena.
Both pre-existing large-scale magnetohydrodynamic (MHD) turbulence and locally 
generated turbulence through plasma instabilities are important for the CR propagation 
in astrophysical media.
The potential role of the resonant instability triggered by CR pressure anisotropy
to regulate the parallel spatial diffusion of low-energy
CRs ($\lesssim 100$ GeV) in the interstellar and intracluster medium of galaxies (ISM and ICM) 
has been showed in previous theoretical works.
This work aims to study the gyroresonance instability via direct numerical simulations, 
in order to access quantitatively the wave-particle scattering rates.
For this we employ a 1D PIC-MHD code to follow the growth 
and saturation of the gyroresonance instability.
We extract
from the simulations the pitch-angle diffusion coefficient $D_{\mu\mu}$ produced by the 
instability during the linear and saturation phases, and a very good agreement (within a factor of 3) 
is found with the values predicted by the quasilinear theory (QLT).
Our results support the applicability of the QLT for modeling 
the scattering of low-energy CRs by the gyroresonance instability in the complex interplay
between this instability and the large-scale MHD turbulence.
\end{abstract}

\begin{keywords}
cosmic ray -- 
{\it (magnetohydrodynamics)} MHD --
turbulence --
plasmas
\end{keywords}



\begingroup
\let\clearpage\relax
\endgroup
\newpage

\graphicspath{{./figs/}}

\section{Introduction}

Basic processes of transport of cosmic rays (CRs) are crucial for 
understanding most of high-energy phenomena, ranging from solar flares (\citealt{YLP08}), 
$\gamma$-ray emission of molecular clouds associated to supernova remnants
(\citealt{nava_gabici_2013}), to remote cosmological 
objects such as $\gamma$-ray bursts (\citealt{zhang_yan_2011}).
The interaction of CRs with magnetohydrodynamic (MHD) turbulence is thought to be the principal 
mechanism to scatter and isotropize CRs (e.g. \citealt{ginzburg_1966, jokipii_1966, wentzel_1969, 
schlickeiser_2002}; \citealt{yan_2015} and ref. therein). 
In addition to the large-scale MHD turbulence, small-scale instability generated perturbations 
also play crucial roles. 
In the context of acceleration in supernova shocks, studies of instabilities have been one 
of the major efforts in the field since the acceleration efficiency is essentially determined 
by the confinement at the shock region and magnetic field amplification (e.g. \citealt{bell_lucek_2001, bell_2004, 
YLS12, caprioli_spitkovsky_2014, brose_etal_2016}).

The diffusive propagation of CRs in the Galaxy, far from sources, is considered to be regulated by 
the interactions with the interstellar medium (ISM) background turbulence. 
However, due to the damping 
of the background turbulence (particularly the fast modes), CRs with energies below $\sim 100$ GeV are 
expected to be mainly influenced by the self-generated instabilities. 
In fact, the small-scale instabilities and large-scale turbulence are not independent of each other. 
First, the instability generated waves can be damped through the interaction with the large-scale 
turbulence (\citealt{yan_lazarian_2002, farmer_goldreich_2004, lazarian_2016}). 
Second, the large-scale compressible turbulence also generate small-scale waves through 
thermal plasma (e.g. \citealt{schekochihin_etal_2005, santos-lima_etal_2014, santos-lima_etal_2016, sironi_narayan_2015}) and CR
resonant instabilities (\citealt{lazarian_beresnyak_2006, yan_lazarian_2011}).

The compression/expansion and shear from the large-scale ISM 
turbulence produce deformations in the local particle pitch angle distribution due to the 
conservation of the first adiabatic invariant. Such anisotropic distributions are subjected to various 
instabilities. 
Waves generated through instabilities enhance the scattering rates of the particles, and their 
distribution then relaxed to the state of marginal state of instability even in the collisionless environment 
(\citealt{schekochihin_cowley_2006, kunz_etal_2014, santos-lima_etal_2016, sironi_narayan_2015, riquelme_etal_2015}). 
In particular, the anisotropy in the CR pressure 
induces a (gyro)resonant instability. 
Unlike the streaming instability (e.g. \citealt{skilling_1970, skilling_1971, skilling_1975, amato_blasi_2009}), this gyroresonance instability does not require the bulk motion of CRs. 
The wave grows at the expense of the free energy from CRs' anisotropy induced by the large-scale turbulent motions. 
In the case that the energy growth rate reaches the turbulence energy cascading rate, turbulence is damped
(\citealt{yan_lazarian_2011}). 
This is one of the feedbacks from CRs on turbulence.

\citet[\citetalias{yan_lazarian_2011} hereafter]{yan_lazarian_2011} proposed an analytical equilibrium model of 
the CR diffusion resulting from their scattering by the gyroresonance instability, based on the quasilinear theory (QLT). They derived the dependence of the diffusion 
coefficients with the parameters of the medium turbulence, taking into accout self-consistently the damping 
effect of the instability on the large-scale compressible cascade. They found this mechanism to be important 
for the CR propagation in collisionless medium, such as the Halo and hot ionized medium of the Galaxy, and 
also in the intracluster medium (ICM) of galaxies.

These previous findings motivate more detailed studies on this scattering mechanism
and its role in the CR propagation in both the ISM and ICM.
The saturation state of the instability as well as the 
accuracy
of the CR scattering rates based on the QLT 
are addressed in this work. 
For this aim, we use direct numerical simulations to
follow the evolution of an initially unstable CR distribution propagating 
in a thermal background plasma. These simulations are based on a hybrid technique which combines Particle-In-Cell (PIC) and MHD
(e.g. \citealt{lucek_bell_2000, bai_etal_2015}).
This first study is restricted to unidirectional propagating modes, parallel to
the mean field (one-dimensional geometry).

This paper is organized in the following way: in \S\ref{sec:theory} we present the basic equations which describe
the relevant CR interactions with the background thermal plasma, the dispersion relation 
for the transverse modes propagating parallel to the magnetic field for a distribution 
of CRs with pressure anisotropy, and the QLT prediction for the CR pitch-angle diffusion coefficient.
In \S\ref{sec:numerical} we describe the numerical methods employed in our numerical simulations, and the 
results are presented in \S\ref{sec:results}. Finally, in \S\ref{sec:summary} we summarize our findings and offer our conclusions.

\section{CR pressure anisotropy gyroresonance instability}\label{sec:theory}

\subsection{CR + thermal plasma description}

The relevant physical phenomena for the CR propagation (considered only as ions here) in the ISM or ICM
take place on scales larger or of the order of the CR kinetic scales, many orders of magnitude above 
the kinetic scales of the 
thermal ions in the medium. Hence for studying the effects of the interactions between 
CRs and the thermal plasma it is natural to
retain the kinetic description only for the CRs, 
employing the 
Vlasov equation for the evolution of their distribution function $f (\mathbf{r,p},t)$:
\begin{equation}
\frac{\partial f}{\partial t} + \mathbf{v} \cdot \frac{\partial f}{\partial \mathbf{r}} + q \left( \mathbf{E} + \frac{1}{c} \mathbf{v \times B} \right) \cdot \frac{\partial f}{\partial \mathbf{p}} = 0,
\end{equation}
where $\mathbf{v} = \mathbf{p}/\gamma m_{cr}$ is the CR velocity, $\gamma = \left(1 - v^2/c^2 \right)^{-1/2}$, 
$c$ is the light speed, $m_{cr}$ is the CR rest mass,
$q$ is the CR charge (elementary charge $e$ in our case), $\mathbf{E}$ and $\mathbf{B}$ are the electric and magnetic fields. 
The mass-dominating thermal plasma is described by the MHD approximation, modified to account for the 
CRs presence:
\begin{equation}
    \frac{\partial \rho}{\partial t} + \nabla \cdot (\rho \mathbf{u}) = 0,
\end{equation}
\begin{multline}
    \rho\left[\frac{\partial\mathbf{u}}{\partial t}+(\mathbf{u} \cdot \nabla)\mathbf{u}\right] + \\
    +\nabla P_{th}
    - \frac{1}{4 \pi}(\nabla \times \mathbf{B})\times\mathbf{B} = 
    - \frac{1}{c} \mathbf{J}_{cr}\times\mathbf{B} 
    - q n_{cr} \mathbf{E},
    \label{eq:Byk15}
\end{multline}
\begin{equation}
    \frac{\partial \mathbf{B}}{\partial t} = 
    - c \nabla \times \mathbf{E},
    \label{eq:Byk14}
\end{equation}
\begin{equation}
    \nabla \times \mathbf{B} = \frac{4 \pi}{c} \left(\mathbf{J}_{th} + \mathbf{J}_{cr} \right), \;\;\; \nabla \cdot \mathbf{B} = 0,
\end{equation}
where $\rho$, $\mathbf{u}$, $P_{th}$, $\mathbf{J}_{th}$ are the density, velocity, thermal pressure, and current density fields
of the thermal plasma, $n_{cr}$ and $\mathbf{J}_{cr}$
are the number and current density of the CRs:
\begin{equation}
n_{cr} (\mathbf{r}, t) = \int d \mathbf{p} \; f(\mathbf{r, v}, t),
\end{equation}
\begin{equation}
\mathbf{J}_{cr} (\mathbf{r}, t) = q \int d \mathbf{p} \; \mathbf{v} f(\mathbf{r, v}, t).
\end{equation}
in the ideal case when the resistivity can be neglected, the electric field is given by:
\begin{equation}
    \mathbf{E} = - \frac{1}{c} \mathbf{u \times B}.
    \label{eq:electric_bykov}
\end{equation}

The above equations (\citealt{bykov_etal_2013}) assume (i) the thermal plasma is non-relativistic, and 
(ii) quasi-neutrality of the full plasma: $q_i n_i + q n_{cr} = e n_e$, where $n_{i,e}$ is the number density 
of the thermal ions/electrons, $q_i$ is the ions charge, and $e$ is the elementary charge. 
For simplicity, we consider the thermal plasma composed only by protons and electrons, and the 
CRs as protons ($q = q_i = e$, $m_{cr} = m_i = m_p$, where $m_p$ is the proton mass). 
The system must be closed by an equation for the evolution of $P_{th}$.

\subsection{Development of CR pressure anisotropy}

The momentum distribution of CRs propagating in a region where 
the local magnetic field intensity $B$ is changing slowly in time 
(compared to the CR gyroperiod) becomes distorted 
as the perpendicular component of the particles momentum $p_{\perp}$ modifies according to the conservation of 
the first adiabatic invariant $\propto p_{\perp}^2/B$ (e.g. \citealt{longair_2011}). In this way an initially isotropic 
distribution function 
$f_0(\mathbf{p}) d \mathbf{p} \propto p^{-2 -\alpha} d \mathbf{p}$ assumes an elliptic 
shape in the momentum space:
\begin{equation}
f(\mathbf{p}) d \mathbf{p} \propto (\xi p_{\perp}^2 + p_{\parallel}^2)^{-1 - \alpha/2} d \mathbf{p},
\end{equation}
where $\parallel,\perp$ refer to the directions parallel/perpendicular to the local magnetic field $\mathbf{B}$.
In general, we can parametrize small deviations from the 
isotropy 
by using an expansion
in the distribution of the cosine of the pitch-angle $\mu = p_{\parallel}/|\mathbf{p}|$:
\begin{equation}
\begin{split}
&f(\mathbf{p}) d \mathbf{p} = N(p) p^{2} dp g(\mu) d\mu d\psi, \\
&g(\mu) \propto 1 + 3 \beta \mu + \frac{\chi}{2} \left( 3 \mu^2 - 1 \right),
\end{split}
\label{eq:gmu}
\end{equation}
where $\psi$ is the gyro-phase and $|\beta|, |\chi| \ll 1$.
The above distribution is obviously restricted to the case of 
uniform anisotropy over all $p$.
In this study we neglect the dipole component $\beta$ (no CR bulk velocity).
In this case, the above distribution function translates in an anisotropy $A$ of the CR pressure components 
$P_{\perp,\parallel}$ defined by
\begin{equation}
A \equiv P_{\perp}/P_{\parallel} - 1,
\end{equation}
where
\begin{equation}
P_{\perp} \equiv \frac{1}{2} \int d \mathbf{p} f(\mathbf{p}) v_{\perp} p_{\perp}, \;\;\;
P_{\parallel} \equiv \int d \mathbf{p} f(\mathbf{p}) v_{\parallel} p_{\parallel}.
\end{equation}
The correspondence between $\chi$ and $A$ is
\begin{equation}
\chi = \frac{5 A}{2 A + 3} = \frac{5}{3} A + \mathcal{O} \left( A^2 \right).
\end{equation}

The analytical model in \citetalias{yan_lazarian_2011} provides the following estimates for $A$:
$\gtrsim 10^{-3}$ for the galactic Halo and $\gtrsim 10^{-4}$ for the hot interstellar medium 
and ICM. 
Although this work is dedicated to study the isolated effects of CR pressure anisotropy,
it should be observed that we expect 
a large scale drift of CRs flowing from the galactic sources to outside the Galaxy,
with this dipole anisotropy (of the order of $\beta$) observed to be $10^{-4}-10^{-3}$ at the Earth \citep{skilling_1970, disciascio_iuppa_2014}.
A systematic study on the combined effect of both kinds of anisotropy is beyond 
the scope of the present work, and will be addressed elsewhere.

\subsection{Parallel propagating transverse modes}

The dispersion relation of the ordinary linear MHD waves can be modified depending on the
CRs distribution function. In fact the free-energy provided by the anisotropic 
pressure of CRs can turn the MHD waves unstable (e.g. \citealt{schlickeiser_2002}). 
We focus here only on the parallel propagating modes, for simplicity and also 
because they have the fastest growth rates,
and therefore should be more important 
for CR scattering.
\citet{bykov_etal_2013} present the linear dispersion relation for the general case 
of a distribution function described by Eq.~\ref{eq:gmu}.
In the absense of CR bulk velocity ($\beta = 0$), the
linear dispersion relation for $k>0$ 
is~\footnote{The solution in Eq.~\ref{eq:ola1} is rigorously derived 
for $\operatorname{Im}(\omega) > 0$. The validity of this expression for 
$\operatorname{Im}(\omega) < 0$ comes from the analytical continuation of the solution.}:
\begin{equation}
    \omega^2=v_A^2\left\{k^2\mp k\frac{4\pi q n_{cr}\chi}{B_0} \int_{p_{\min}}^{p_{\max}}\sigma(p, k)N(p)p^2dp \right\},
    \label{eq:ola1}
\end{equation}
where 
we assume $|\omega| \ll |\Omega|$, $\Omega = \Omega_0/\gamma$ 
is the CR cyclotron frequency with $\Omega_0 = q B_0/m_{cr} c$,
$v_A=B_0/\sqrt{4\pi\rho}$ is the Alfv\'{e}n speed, $p_{\min}$ and $p_{\max}$ are the minimum and maximum 
values of the CR momentum distribution $N(p)$, and
\begin{equation}
    \sigma(p,k)=\frac{3}{4}\int_{-1}^1\frac{(1-\mu^2)\mu}{1\mp k v \mu/\Omega}d\mu,
    \label{eq:Byk23}
\end{equation}
where $\mp$ correspond to the Left and Right circular polarization, respectively.
We adopt the definition of polarization used in \citet{stix_1962} and \citet{gary_1993}: the waves with Left polarization 
rotate in the same sense as protons, irrespective to the wave propagation direction, while the waves with Right polarization 
rotate with the same sense as electrons.

According to the dispersion relation given by Eq.~\ref{eq:ola1}, for small anisotropies ($|\chi| \ll 1$) the real ($\omega_r$) and imaginary ($\Gamma$) 
parts of the frequency are: 
\begin{equation}
|\omega_r (k)| = v_A k + \mathcal{O} (\chi),
\label{eq:omega_simplified}
\end{equation}
\begin{multline}
\Gamma^{L,R} (k) = \mp v_A \frac{2\pi q n_{cr}\chi}{B_0} \operatorname{Im} \left\{ \int_{p_{\min}}^{p_{\max}}\sigma(p, k)N(p)p^2dp \right\} + \\
+ \mathcal{O} (\chi^2).
\label{eq:gamma_int}
\end{multline}

Using the analytical solution in \citet{bykov_etal_2013}, the imaginary part in $\sigma(p,k)$ (Eq.~\ref{eq:Byk23}) comes from the pole 
(resonant) contribution and 
is given by:
\begin{multline}
\operatorname{Im} \left\{ \sigma(p,k) \right\} = - \frac{3 \pi}{4} \left[ \left( \frac{\Omega_0 m_{cr}}{k p} \right)^{2} - \left( \frac{\Omega_0 m_{cr}}{k p} \right)^{4} \right] \times
\\ \times \mathcal{H} \left( p - m_{cr} \Omega_0/k \right)
\end{multline}
where $\mathcal{H}(x)$ is the Heaviside step function. 
Assuming the following power law for the momentum distribution:
\begin{equation}
N(p) = \frac{ (1 - \alpha) }{ \left( p_{\max}^{1 - \alpha} - p_{\min}^{1 - \alpha} \right) } p^{-2 - \alpha}
\label{eq:distr_p}
\end{equation}
for $p_{\min} < p < p_{\max}$ and zero otherwise, the integration in Eq.~\ref{eq:gamma_int} gives:
\begin{multline}
  \operatorname{Im} \left\{ \int_{p_{\min}}^{p_{\max}}\sigma(p, k)N(p)p^2dp \right\} (k) \approx 
  \\ \approx - \frac{3 \pi}{4} \frac{ (\alpha - 1) }{(\alpha + 1)} \times \left \{
  \begin{aligned}
    &0, && \text{if}\ k r_{\min} < p_{\min}/p_{\max} \\
    &( k r_{\min} )^{\alpha - 1}, && \text{if}\ p_{\min}/p_{\max} < k r_{\min} < 1 \\
    &( k r_{\min} )^{- 2}, && \text{if}\ k r_{\min} > 1
  \end{aligned} \right.
\end{multline} 
where $r_{\min} \equiv p_{\min}/m_{cr} \Omega_0$ is the Larmor radius for the lower energy particles (in the limit of zero pitch-angle), and assuming $\alpha > 2$, $p_{\max}/p_{\min} \gg 1$.
Now we can rewrite Eq.~\ref{eq:gamma_int} as:
\begin{multline}
  \Gamma^{L,R} (k) \approx 
  \pm \frac{5 \pi}{8} \frac{ (\alpha - 1) }{(\alpha + 1)}
  \left( \frac{c}{v_A} \right) \Omega_0 
  \left( \frac{n_{cr}}{n_i} \right)
  A 
  \times \\
  \times \left \{
  \begin{aligned}
    &0, && \text{if}\ k r_{\min} < p_{\min}/p_{\max} \\
    &( k r_{\min} )^{\alpha - 1}, && \text{if}\ p_{\min}/p_{\max} < k r_{\min} < 1 \\
    &( k r_{\min} )^{- 2}, && \text{if}\ k r_{\min} > 1 .
  \end{aligned} \right.
\label{eq:gamma_simplified}
\end{multline} 

\subsection{Quasilinear instability evolution}

The parallel propagating Alfv\'{e}n waves with polarization $L$ (if $A>0$) or $R$ (if $A<0$)
grow exponentially and scatter resonantly the CRs. This scattering 
can be described by a pitch-angle diffusion coefficient in the Fokker-Planck equation for the evolution of the 
ensemble averaged distribution function $\langle f(\mathbf{r, p}, t) \rangle$:
\begin{equation}
\left( \frac{\partial \langle f \rangle}{\partial t} \right)_{scatt} = 
\frac{\partial}{\partial \mu} \left( D_{\mu\mu} \frac{\partial \langle f \rangle}{\partial \mu} \right).
\end{equation}

The dependence of the diffusion coefficient $D_{\mu\mu}$ with the electromagnetic waves properties (dispersion and amplitude)
can be derived analytically under the quasilinear theory, valid for small amplitude random-phased waves.
In Appendix~\ref{sec:appendix} we present a derivation 
of $D_{\mu\mu}$ using a procedure equivalent to the quasilinear theory. Under the set of hypothesis 
detailed in the Appendix and observing that 
the real frequency of the waves $L$ and $R$ are almost identical, i.e. $\omega_{r,n}^{L}(k) \approx \omega_{r,n}^{R}(k) = \omega_r (k)$
($n=f,b$ for forward or backward propagation directions, respectively), we obtain:
\begin{multline}
D^{QL}_{\mu\mu} (\delta t) = \frac{\Omega^2 (1 - \mu^2)}{2} \int_{0}^{\infty} d k 
\frac{|\mathbf{B}(k)|^2}{B_0^2} \times \\
\times \bigl\{ [1 + \sigma_H(k)] \mathcal{R} (v \mu k - \omega_{r} + \Omega, \delta t) + \bigr. \\
\bigl. [1 - \sigma_H(k)] \mathcal{R} (v \mu k - \omega_{r} - \Omega, \delta t) \bigr\}
,
\end{multline}
where the
resonance function $\mathcal{R}$ is given by Eq.~\ref{eq:res_func}, $\sigma_H(k)$ is the total helicity of the waves component 
of $\mathbf{B}(k)$. $\sigma_H = - 1$ for 
waves with $L$ polarization and forward propagation or $R$ polarization and backward propagation; 
$\sigma_H = +1$ for $L$ polarization and backward propagation or $R$ polarization and forward propagation.
The time interval $\delta t$ is understood as the time interval under which the diffusion process is considered, 
and it is much shorter than the timescale of change of the distribution function due to the scattering itself
(see Appendix~\ref{sec:appendix}). 
Therefore, the above definition of the diffusion coefficient is valid for describing the evolution
of distribution functions averaged in a time 
interval $\sim \delta t$.
Because the system we are focusing in this study is statistically homogeneous and the theoretical 
$D^{QL}_{\mu\mu}$ does not depend on the 
gyro-phase, we can consider the average of the Fokker-Planck equation in space 
and gyro-phase.

The effect of $D_{\mu\mu}$ is to reduce gradually the anisotropies in $\mu$ of the distribution function: the source of free-energy of the instability. 
Therefore after the linear phase,
the initial distribution function of the CRs 
will evolve reducing the $\mu$ anisotropy, then reducing the instability growth rate. During this last saturation 
phase, the distribution function will gradually relax to a stable distribution (isotropic).
The diffusion timescale depends on $\Omega$, or on the particle energy. This means that the anisotropy distribution over $p$ 
does not evolve uniformly, i.e. $A = A(p)$ after the linear phase.

The CR mean-free-path along the field lines (from the scattering 
of particles along this direction) is related to $D_{\mu\mu}$ via:
\begin{equation}
\lambda_{\parallel} 
= \frac{3}{4} v \int_{-1}^{1} d\mu \frac{( 1 - \mu^2 )}{\nu_{scatt}}
= \frac{3}{8} v \int_{-1}^{1} d\mu \frac{( 1 - \mu^2 )^2}{D_{\mu \mu}},
\end{equation}
where $\nu_{scatt} \equiv 2 D_{\mu\mu}/(1 - \mu^2)$ is the CR scattering rate.

\section{Numerical methods}\label{sec:numerical}

In order to study the evolution of the CR pressure anisotropy instability described by 
the MHD + CR kinetic equations (see \S2.1), we use an hybrid MHD-particle code which 
represents the MHD fields ($\rho$, $\mathbf{u}$, $P_{th}$, $\mathbf{B}$)
in cells defined by a grid over the 
simulation domain, while the CR distribution function is sampled by a collection of macro-particles 
which orbits are directly solved. 
The CR macroscopic fields
needed for the evolution of the MHD variables ($n_{cr}, \mathbf{J}_{cr}, \mathbf{u}_{cr}$)
are calculated by a process of deposition of the macro-particles on the grid cells (Particle-In-Cell method).
This MHD-PIC coupling is described in detail in \citet{bai_etal_2015}.

The grid cell variables are evolved by the following set of equations:
\begin{equation}
\frac{\partial \rho}{\partial t} + \nabla \cdot (\rho \mathbf{u}) = 0,
\end{equation}
\begin{equation}
\frac{\partial}{\partial t} (\rho \mathbf{u}) + \nabla \cdot \left\{ \rho \mathbf{u u} + \left[ P_{th} + \frac{B^2}{8 \pi} \right] \mathbf{I} - \frac{\mathbf{B B}}{4 \pi} \right\} = - \mathbf{F}_{cr},
\end{equation}
\begin{multline}
\frac{\partial \epsilon}{\partial t} + \nabla \cdot \left\{ \left[ \epsilon + P_{th} + \frac{B^2}{8 \pi} \right] \mathbf{u} - \frac{\mathbf{\left( u \cdot B \right) B}}{4 \pi} \right. + \\
\left. + \frac{c}{4 \pi} \mathbf{(E - E_0) \times B} \right\} = - \mathbf{u}_{cr} \cdot \mathbf{F}_{cr},
\end{multline}
\begin{equation}
\frac{\partial \mathbf{B}}{\partial t} + c \nabla \times \mathbf{E} = 0,
\end{equation}
where $\mathbf{I}$ is the unitary dyadic tensor, $\epsilon = \rho u^2/2 + B^2/8\pi + P_{th}/(\gamma_{th} - 1)$ is the total energy density with $\gamma_{th} = 5/3$ the adiabatic index of the thermal gas, 
$\mathbf{E}_0$ and $\mathbf{E}$ are the electric fields given by
\begin{equation}
\mathbf{E}_0 = - \frac{1}{c} \mathbf{u} \times \mathbf{B},
\end{equation}
\begin{equation}
\mathbf{E} = \mathbf{E}_0 - \frac{1}{c} \frac{n_{cr}}{n_e} (\mathbf{u}_{cr} - \mathbf{u}) \times \mathbf{B},
\label{eq:electric_hall}
\end{equation}
where $n_e = n_{cr} + n_i$ is as before the electrons number density, and $\mathbf{F}_{cr}$ is the force density felt by the CRs:
\begin{equation}
\mathbf{F}_{cr} = q n_{cr} \mathbf{E} + \frac{1}{c} \mathbf{J}_{cr} \times \mathbf{B}.
\end{equation}

The orbits of the macro-particles are evolved using the
the Lorentz force with the electromagnetic fields interpolated from the grid values:
\begin{equation}
\frac{d \mathbf{p}_j}{dt} = q \mathbf{E} + \frac{q}{c} \mathbf{v}_j \times \mathbf{B}
\end{equation}
where $\mathbf{p}_j$ and $\mathbf{v}_j$ are the momentum and velocity of the particle $j$.

It should be pointed out that the electric field $\mathbf{E}$ (Eq.~\ref{eq:electric_hall}) 
differs from $\mathbf{E}_0$ by the inclusion of the CR Hall effect 
which is not taken into 
account in Eq.~\ref{eq:electric_bykov}. However, the effects of this term (of the order of $n_{cr}/n_i$ for $n_{cr} \ll n_i$)
can be considered negligible if $\mathbf{u}_{cr} \sim \mathbf{u}$, which is the case for the 
transverse Alfv\'{e}n modes we are interested in. 

The above equations are solved in a periodic, cartesian, one-dimensional domain.
The MHD equations are discretized using an conservative formulation, the fluxes are calculated using 
the HLLD solver (adapted from the ATHENA code, \citealt{stone_etal_2008}) and linear interpolation. 
The particles are evolved using the relativistic Boris pusher (adapted from the 
SKELETON particle-in-cell codes, \citealt{decyk_1995, decyk_2007})
with a leap-frog scheme, and first-order weighting for the particles deposit on the grid. 
The full time integration (MHD + particles) is performed using Runge-Kutta of second order.
We verified the second order convergence of our code implementation against several linear test 
problems, as MHD waves with cold and not-cold CR distributions (with different relative densities $n_{cr}/n_i$) and 
the non-resonant Bell instability.

\section{Results}\label{sec:results}

\subsection{Simulations parameters}

The initial conditions of the performed simulations consist of homogeneous 
MHD fields with null-velocity and mean magnetic field parallel to the axis of the simulation grid,
and a distribution of CRs  
with anisotropy $A_0$
(distribution $g(\mu)d\mu$ given by Eq.~\ref{eq:gmu}), superimposed 
by a flat spectrum of circularly polarized waves with magnetic field amplitude $|B_{k0}|$.

The momentum distribution of CRs has power-law index $\alpha = 2.8$, with 
$\gamma(p_{\min}) = 2$ and $p_{\max}/p_{\min}=10$ (distribution $N(p)dp$ given by Eq.~\ref{eq:distr_p}). 
We fix the ratio $v_{A}/c = 10^{-2}$.
We vary the parameters $A_0$ and $n_{cr}/n_i$. 

Tables~\ref{tab:runs_param_comm} and ~\ref{tab:runs_param} summarizes the parameters used for the 
simulations: $n_{cr}/n_i$, $v_A/c$, ratio between CR kinetic and magnetic energy density $\beta_{cr} \equiv W_{K,cr}/W_B$, 
ratio between thermal and magnetic pressures $P_{th}/P_B$, 
domain size $L$, 
grid resolution NX, number of particles per grid cell NP/NX,
$A_0$, $(|B_{k0}|/B_0)^2$, polarization $P$, and total helicity $\sigma_H$.

\begin{table*}
\caption{Setup parameters for all the MHD-PIC simulations}
\begin{tabular}{c c c c c c c c c c c}
\hline
$v_{A}/c$ & $P_{th} / P_B$ & $\gamma (p_{\min})$ & $p_{\max}/p_{\min}$ & $\alpha$ & $L \Omega_0 / c$ \\
\hline
$10^{-2}$ & $2$ & $2$ & $10$ & $2.8$ & $500$ \\
\hline
\end{tabular}
\label{tab:runs_param_comm}
\end{table*}
\begin{table*}
\caption{Model parameters}
\begin{tabular}{c c c c c c c c c c c c}
\hline
run & $A_0$ & $n_{cr}/n_i$ & $\beta_{cr}$ & $|\Gamma_{\max}|/\Omega_0$ & $|B_{k0}|^2/B_0^2$ & $P$ & $\sigma_H$ & $t_{end} \Omega_0$ & NX & NP/NX \\ 
\hline
$d1$ & $-0.3$ & $10^{-4}$ & $5.2$ & $1.4 \times 10^{-3}$ & $5 \times 10^{-4}$ & $R(+1)$ & $+1$ & $10^{3}$ & $4096$ & $4096$ \\
$d2$ & $-0.3$ & $10^{-4}$ & $5.2$ & $1.4 \times 10^{-3}$ & $5 \times 10^{-4}$ & $L(-1)$ & $-1$ & $10^{3}$ & $4096$ & $4096$ \\
\hline
$l1$ & $+0.1$ & $10^{-4}$ & $5.2$ & $3 \times 10^{-4}$ & $10^{-12}$ & $L(-1)$ & $0$ & $2 \times 10^{3}$ & $4096$ & $4096$ \\
$l2$ & $+0.2$ & $10^{-4}$ & $5.2$ & $6 \times 10^{-4}$ & $10^{-12}$ & $L(-1)$ & $0$ & $2 \times 10^{3}$ & $4096$ & $4096$ \\
$l3$ & $+0.3$ & $10^{-4}$ & $5.2$ & $9 \times 10^{-4}$ & $10^{-12}$ & $L(-1)$ & $0$ & $2 \times 10^{3}$ & $4096$ & $4096$ \\
$l4$ & $+0.2$ & $2.5 \times 10^{-5}$ & $1.3$ & $2 \times 10^{-4}$ & $10^{-12}$ & $L(-1)$ & $0$ & $2 \times 10^{3}$ & $4096$ & $4096$ \\
$l5$ & $+0.2$ & $4 \times 10^{-4}$ & $20.8$ & $2.4 \times 10^{-3}$ & $10^{-12}$ & $L(-1)$ & $0$ & $2 \times 10^{3}$ & $4096$ & $4096$ \\
\hline
$s1$ & $+0.1$ & $10^{-4}$ & $5.2$ & $3 \times 10^{-4}$ & $2 \times 10^{-12}$ & $R,L$ & $0$ & $2 \times 10^{5}$ & $4096$ & $1024$ \\
$s2$ & $0$ & $10^{-4}$ & $5.2$ & $0$ & $2 \times 10^{-12}$ & $R,L$ & $0$ & $2 \times 10^{5}$ & $4096$ & $1024$ \\
$s3$ & $-0.1$ & $10^{-4}$ & $5.2$ & $4 \times 10^{-4}$ & $2 \times 10^{-12}$ & $R,L$ & $0$ & $2 \times 10^{5}$ & $4096$ & $1024$ \\
$s4$ & $-0.2$ & $10^{-4}$ & $5.2$ & $8 \times 10^{-4}$ & $2 \times 10^{-12}$ & $R,L$ & $0$ & $2 \times 10^{5}$ & $4096$ & $1024$ \\
$s5$ & $-0.3$ & $10^{-4}$ & $5.2$ & $1.4 \times 10^{-3}$ & $2 \times 10^{-12}$ & $R,L$ & $0$ & $2 \times 10^{5}$ & $4096$ & $1024$ \\
$s6$ & $0$ & $10^{-4}$ & $5.2$ & $0$ & $2 \times 10^{-12}$ & $R,L$ & $0$ & $10^{5}$ & $4096$ & $512$ \\
$s7$ & $0$ & $10^{-4}$ & $5.2$ & $0$ & $2 \times 10^{-12}$ & $R,L$ & $0$ & $10^{5}$ & $4096$ & $2048$ \\
\hline
\end{tabular}
\label{tab:runs_param}
\end{table*}

The choice of parameters ($v_A/c$, $n_{cr}/n_i$) is motivated by the conditions in the Galactic Halo 
and hot ionized medium, estimated in \citetalias{yan_lazarian_2011}.
The high values of $\beta_{cr}$ and $A_0$ (compared to the estimates by \citetalias{yan_lazarian_2011}) are chosen in order to 
maximize the growth rate 
of the fluctuations in the resonant energy interval, so that the waves amplified by the instability achieve
higher amplitudes than the magnetic fluctuations (noise) caused by the limited number of macro-particles which sample 
the CR distribution function.

\subsection{Linear phase of instability}

\begin{figure}
\centering 
\includegraphics[width=\columnwidth]{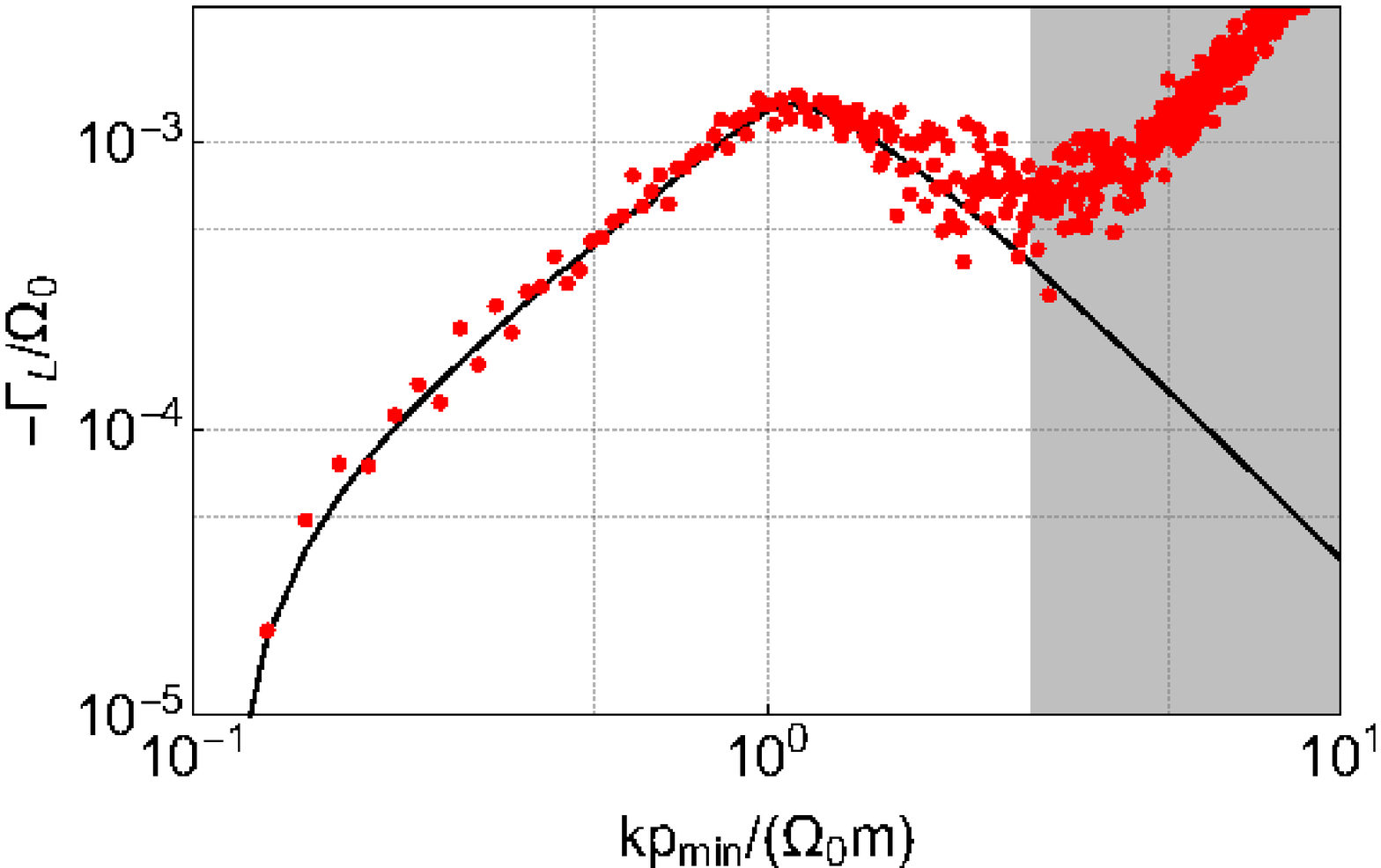} \\
\includegraphics[width=\columnwidth]{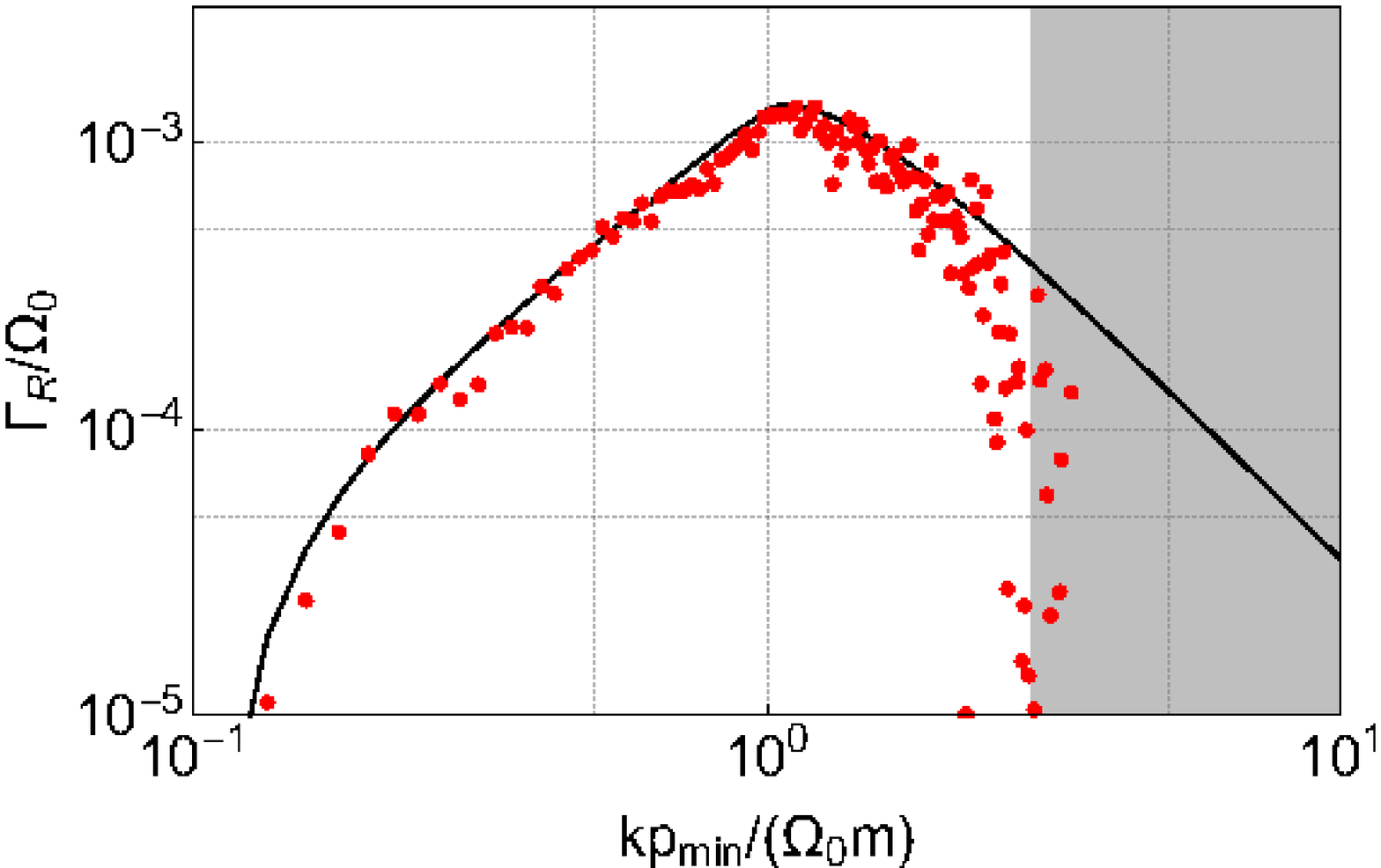}
\caption{Wave growth/damping rate for models with initial anisotropy $A_0 = -0.3$ 
(models $d1$ and $d2$ in Table~\ref{tab:runs_param}). {\it Upper plot:} damping rate of waves with polarization $L$. 
{\it Lower plot:} growth rate of waves with polarization $R$.
The theoretical value given by Eq.~\ref{eq:ola1} is shown for comparison (solid black line).
The gray area comprehends the wavelengths $\le 32$ grid cells, where the growth/damping rate is dominated by the numerical dissipation. }
\label{fig:growth_rate}
\end{figure}

\begin{figure}
\centering 
\includegraphics[width=\columnwidth]{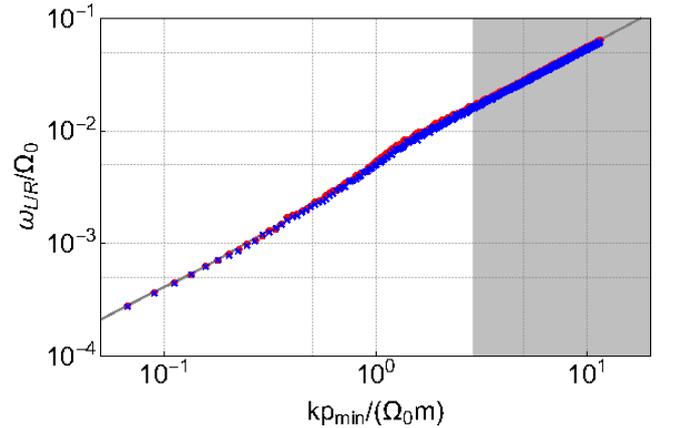}
\caption{Real frequency of waves for models with initial anisotropy $A_0 = -0.3$ 
(models $d1$ and $d2$ in Table~\ref{tab:runs_param}). 
{\it {\color{red} Red points:}} waves with polarization $L$; {\it {\color{blue} blue crosses:}} waves with polarization $R$.
The theoretical value given by Eq.~\ref{eq:ola1} is shown for comparison (solid black line).
The gray area comprehends the wavelengths $\le 32$ grid cells, where the growth/damping rate is dominated by the numerical dissipation.}
\label{fig:phase_speed}
\end{figure}

Figures~\ref{fig:growth_rate} and~\ref{fig:phase_speed} show the dispersion relation extracted from the runs $d1$ and $d2$ 
($A_0 = -0.3$, waves with polarization $R$ and $L$, respectively).
Each point in the figures is calculated by fitting the time-serie of 
one Fourier component of the magnetic field, between
$t=0$ and $t=t_{end}$ (all the points of the same color/style in each plot of Figures~\ref{fig:growth_rate} 
and~\ref{fig:phase_speed}
are extracted from one single run).
In this way we determine for each wavenumber $k$ the phase, the phase speed,
and the growth/damping rate.
Figure~\ref{fig:growth_rate} shows the growth/damping rate $\Gamma(k)$, while Figure~\ref{fig:phase_speed} 
shows the real frequency $\omega_{r}(k)$ of the waves spectrum.
The analytical solution from the dispersion relation (Eq.~\ref{eq:ola1}) is shown for comparison.
The fitted values agree quite well 
with the theoretical values, except for large wavenumbers (inside the gray area), for which the numerical dissipation $\propto k^{2}$ 
dominates $\Gamma(k)$.
Such agreement is observed also in the other runs, 
with the agreement in $\Gamma(k)$ better for the simulations with larger values of $|\Gamma_{\max}|$. For anisotropies 
smaller than $|A_0| = 0.1$ (and $v_A/c = 10^{-2}$, $n_{cr}/n_i = 10^{-4}$), for the same resolution and number of particles, 
the quality of the fitted values for $\Gamma(k)$ decays faster, due to the noise caused by the limited number of particles.


\begin{figure*}
\begin{tabular}{c c} 
\includegraphics[width=\columnwidth]{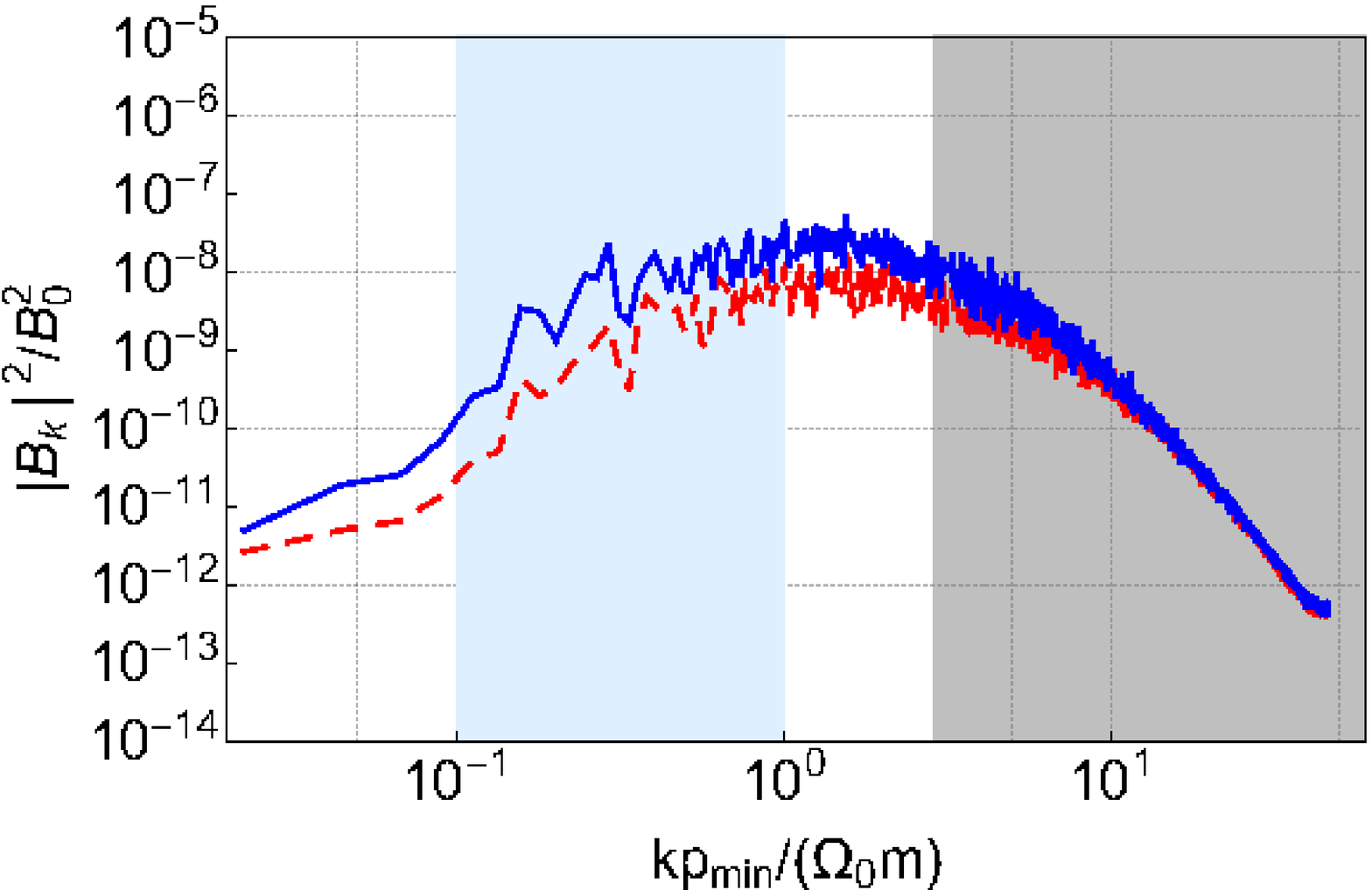} &
\includegraphics[width=\columnwidth]{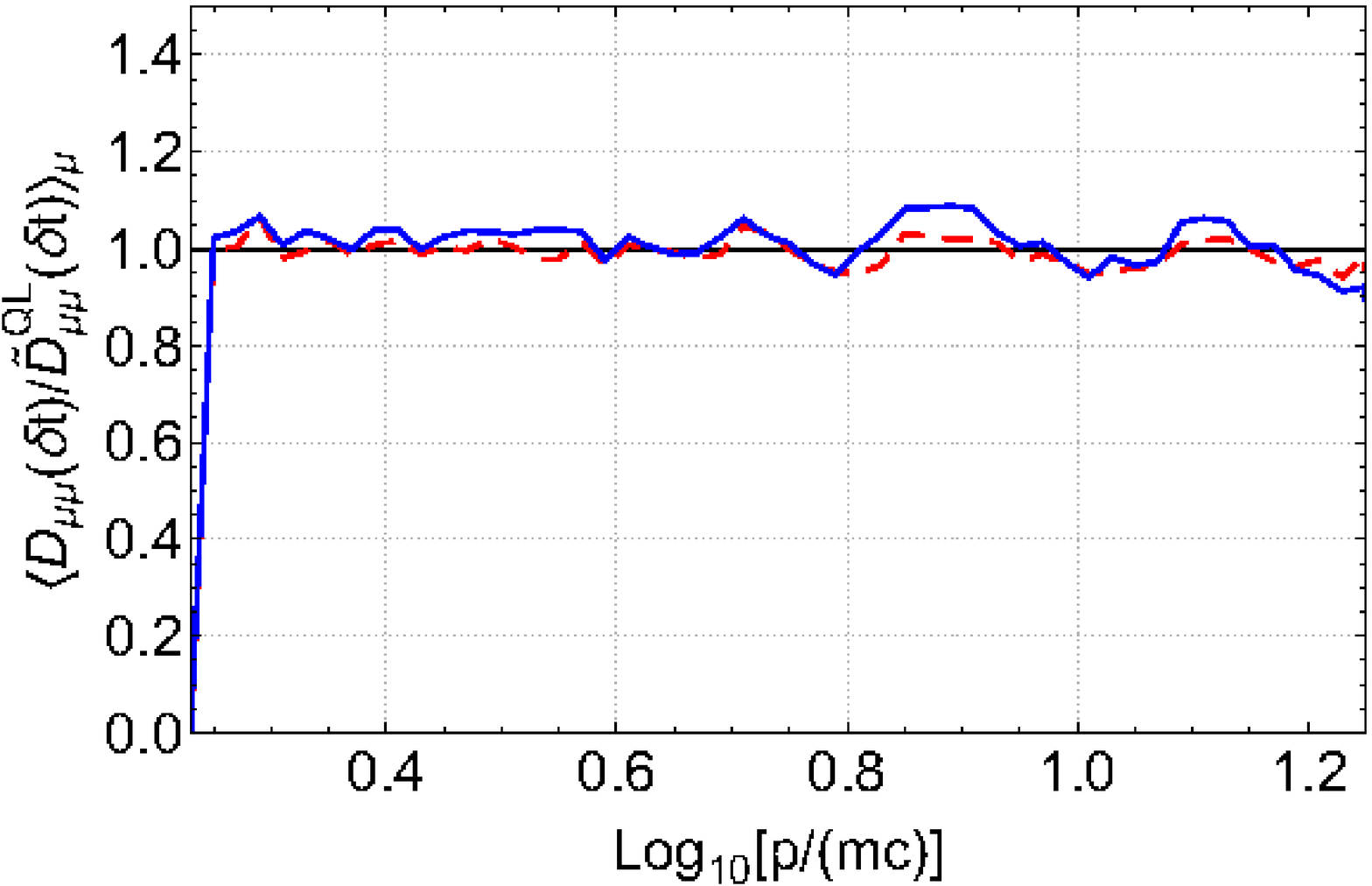} \\
\includegraphics[width=\columnwidth]{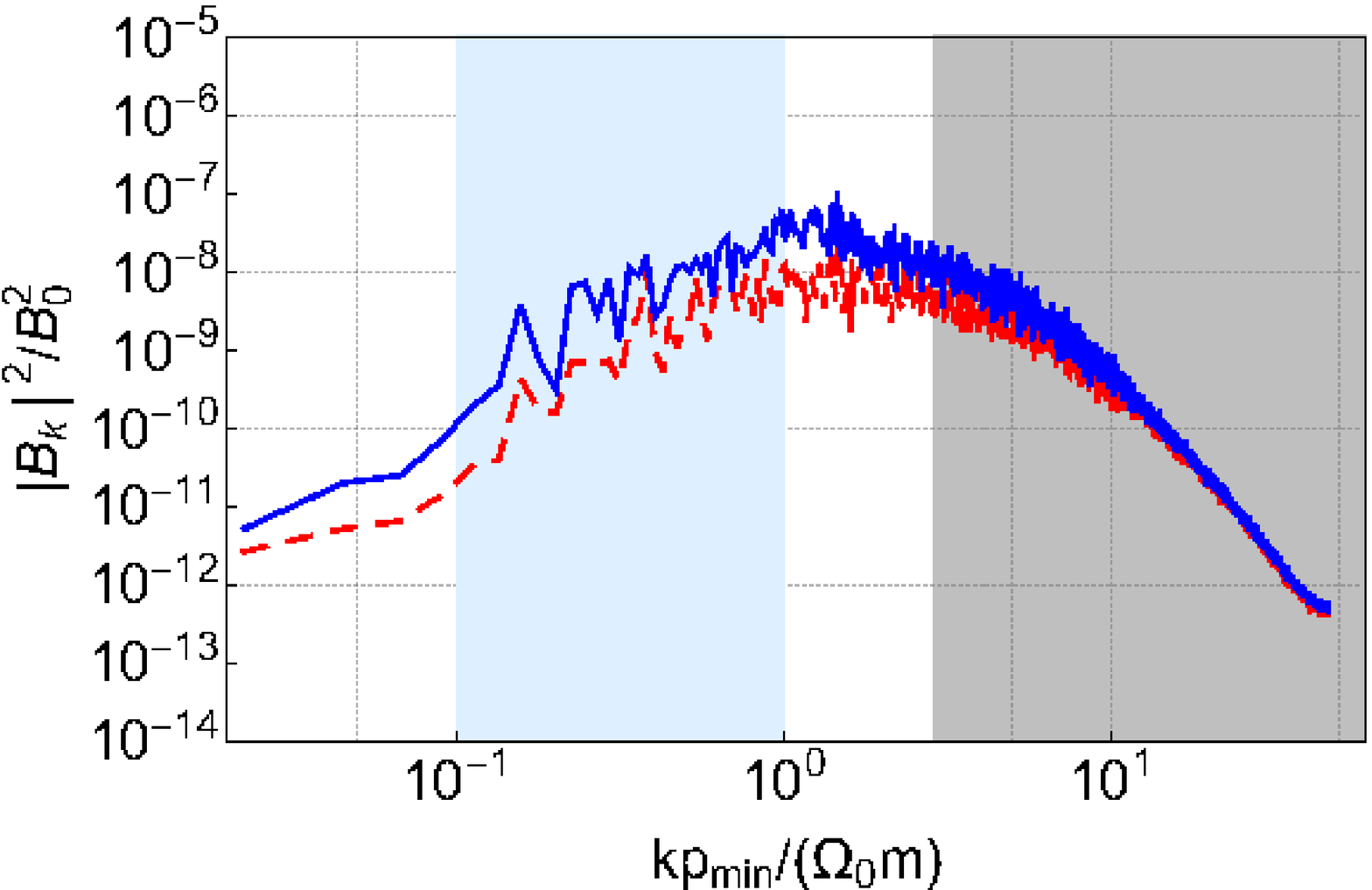} &
\includegraphics[width=\columnwidth]{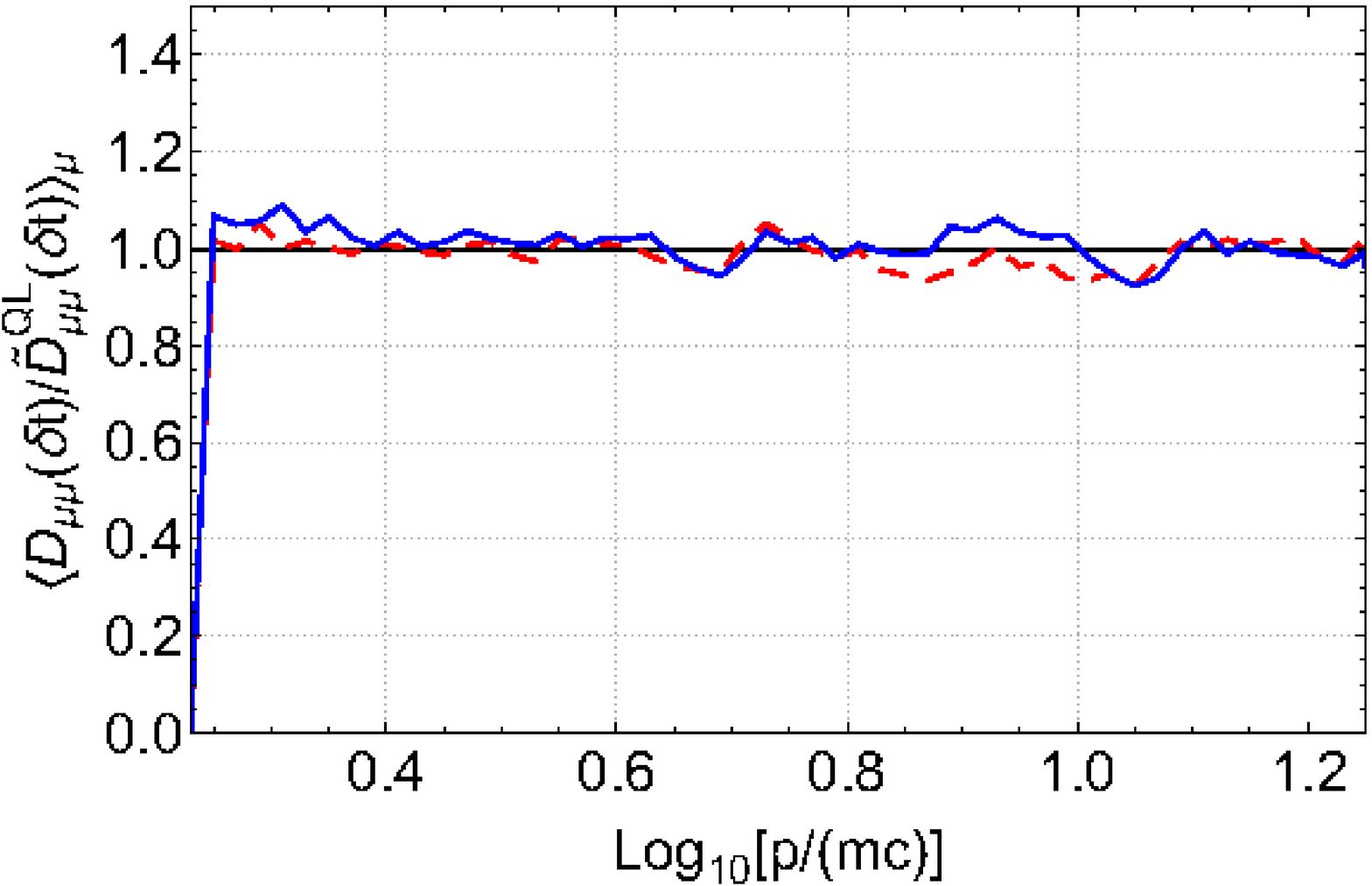} \\
\includegraphics[width=\columnwidth]{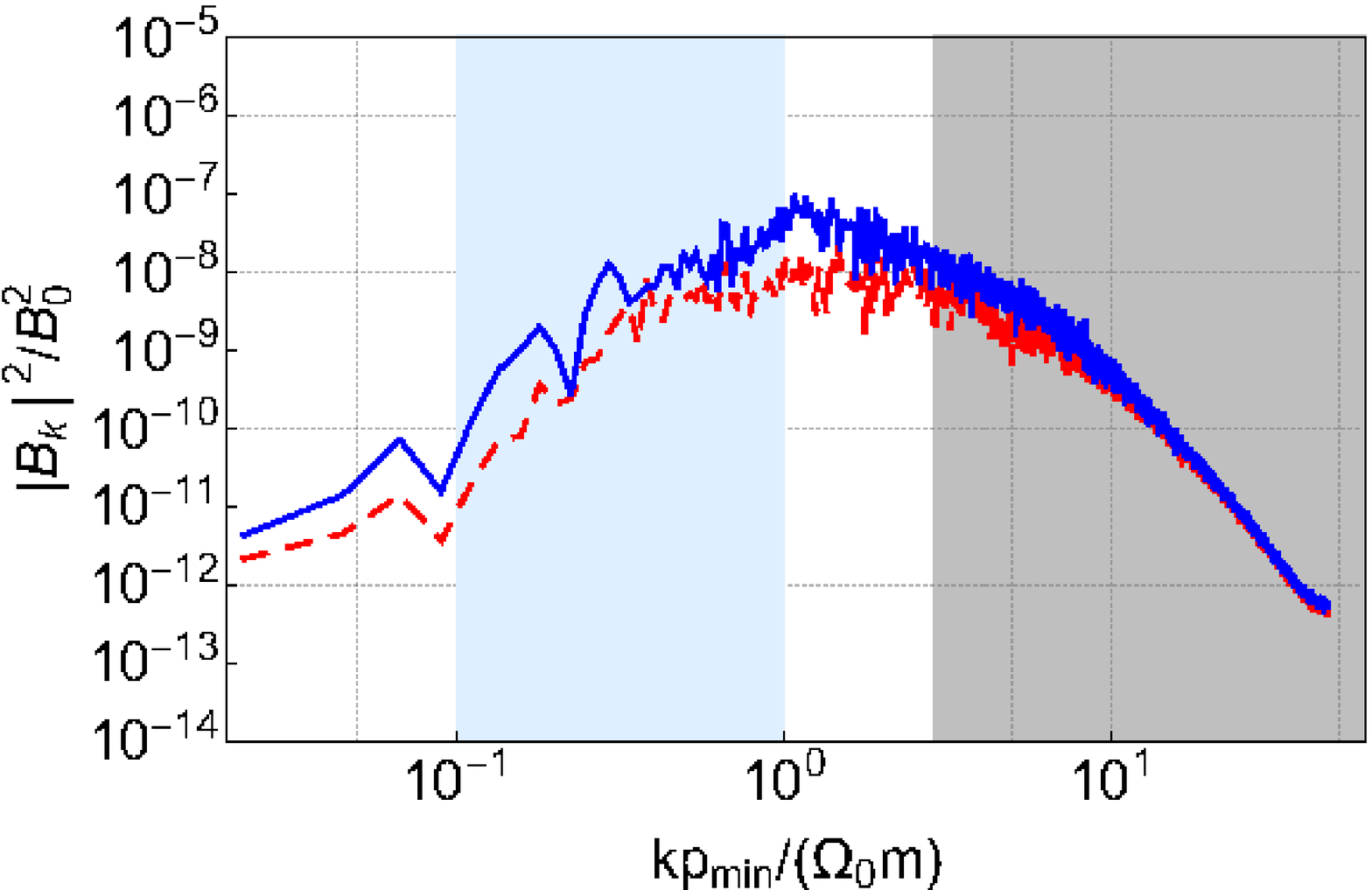} &
\includegraphics[width=\columnwidth]{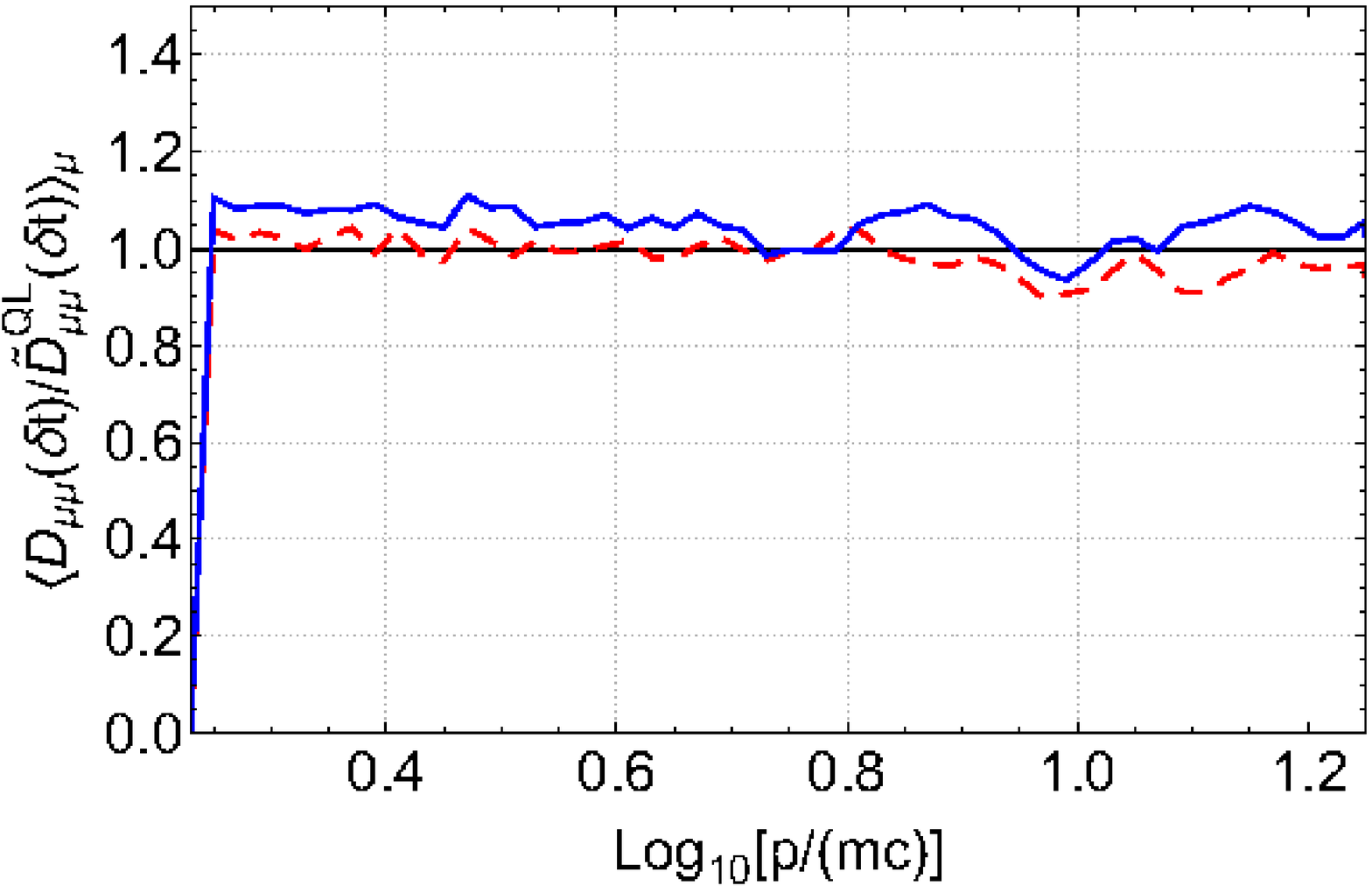}
\end{tabular}
\caption{Normalized power spectrum of magnetic field $|B(k)|^2/B_0^2$ ({\it left column}) and 
$\langle D_{\mu\mu}(\delta t) / \widetilde{D}_{\mu\mu}^{QL} (\delta t) \rangle_{\mu}$ ratio ({\it right column}) for models 
starting with different anisotropies $A_0$: $+0.1$ (top row), $+0.2$ (middle row), $+0.3$ (lower row), corresponding to models $l1$-$l3$ in 
Table~\ref{tab:runs_param}. Lines with different colors indicate different times: 
$t \Omega_0 = 10^3$ ({\color{red} red dashed line}), 
$t \Omega_0 = 2 \times 10^3$ ({\color{blue} blue solid line}).
In the power spectrum plots, the region where the numerical dissipation dominates 
are gray-shaded (wavelengths $\le 32$ grid cells); the blue shaded region represents the interval $\Omega_0 m_{cr} / p_{\max} < k < \Omega_0 m_{cr} / p_{\min}$.
}
\label{fig:short_anis0}
\end{figure*}

We present in the left column of Figure~\ref{fig:short_anis0} the magnetic field power spectrum of models 
with different initial anisotropy $A_0 = +0.1, +0.2, +0.3$ (models $l1$-$l3$) at different times 
during the linear phase of the instability, when the distribution of particles is still almost identical to the 
initial one and the magnetic energy in the waves increase exponentially. The blue shaded region shows the wave numbers for which the growth rate 
(for waves with polarization $L$) is a power law with index related to the CR momentum distribution 
power-law index. The power spectrum in this region increases in amplitude but keeping the slope during the 
measured times, this increase is faster for the higher absolute initial anisotropy. For wavenumbers smaller than those in the blue-shaded region,
the amplitude remains nearly constant, as expected for the zero growth rate in this region. 

Using this magnetic energy spectrum, we estimate the lower limit of the pitch-angle diffusion coefficient 
$\widetilde{D}^{QL}_{\mu\mu} (\delta t)$ provided by the 
quasilinear theory (Eq.~\ref{eq:dmumu_min} in Appendix):
\begin{multline}
\widetilde{D}^{QL}_{\mu\mu} (t, \mu, p, \delta t) \equiv \frac{\Omega^2 (1 - \mu^2)}{2} \int_{0}^{\infty} d k 
\frac{\overline{ |\mathbf{B}(k, t)|^2 }}{B_0^2} \times \\
\times \left\{ [1 + \overline{\sigma_H(k,t)}] \frac{\sin \left[ \left(v \mu k + \Omega \right) \delta t \right]}{\left(v \mu k + \Omega \right)} + \right. \\
\left. + [1 - \overline{\sigma_H(k,t)}] \frac{\sin \left[ \left(v \mu k - \Omega \right) \delta t \right]}{\left(v \mu k - \Omega \right)} \right\},
\end{multline}
where the overbar means an average in time between $t-\delta t$ and $t$; here 
we neglect the real frequency of the waves $\omega_{r}(k)$ ($\ll \Omega$) in the resonance function. 
The helicity spectrum $\sigma_H(k,t)$ is calculated from the transverse electric field spectrum
(\citealt{gary_1993}):
\begin{equation}
\sigma_H (k,t) = - 2 \frac{k}{|k|} \frac{\operatorname{Im} \left\{ E_y (k,t) E^*_z (k,t) \right\}}{|\mathbf{E}(k, t)|^2}.
\end{equation}

We compare the quasilinear estimative with the directly measured diffusion coefficient
\begin{equation}
    D_{\mu\mu}(t, \mu, p, \delta t) = \left\langle
    \frac{[\mu(t) - \mu(t - \delta t)]^2}{2 \delta t}
    \right\rangle ,
\end{equation}
where $\mu(t - \delta t)$ and $\mu(t)$ are cosine of the pitch angle for the same particle at two consecutive times
separated by $\delta t$ (see for example \citealt{xu_yan_2013, weidl_etal_2015, cohet_marcowith_2016}). 
The average $\left\langle \cdot \right\rangle$ is taken over all the particles in the simulation with momentum and 
pitch-angle cosine in a tiny interval $[p, p + \Delta p)$, $[\mu, \mu + \Delta \mu)$.

The distribution in momentum of the ratio 
$\langle D_{\mu\mu}(\delta t)/\widetilde{D}^{QL}_{\mu\mu}(\delta t) \rangle_{\mu}$ 
(averaged over all the values of $\mu$, for $|\mu| \le 0.95$) 
are shown in the right column of Figure~\ref{fig:short_anis0}, 
for the same models and times as for the energy spectra showed in the left column.
We employ a time interval $\delta t \Omega_0 = 10^3$. For the fastest growing mode between these 
runs (model $l5$ in Table~\ref{tab:runs_param}) we have $\delta t \Gamma_{\max} \approx 2.4$.
For all models  the ratio $\langle D_{\mu\mu}(\delta t)/\widetilde{D}_{\mu\mu}(\delta t) \rangle_{\mu}$ 
is very close to one.

\begin{figure*}
\begin{tabular}{c c} 
\includegraphics[width=\columnwidth]{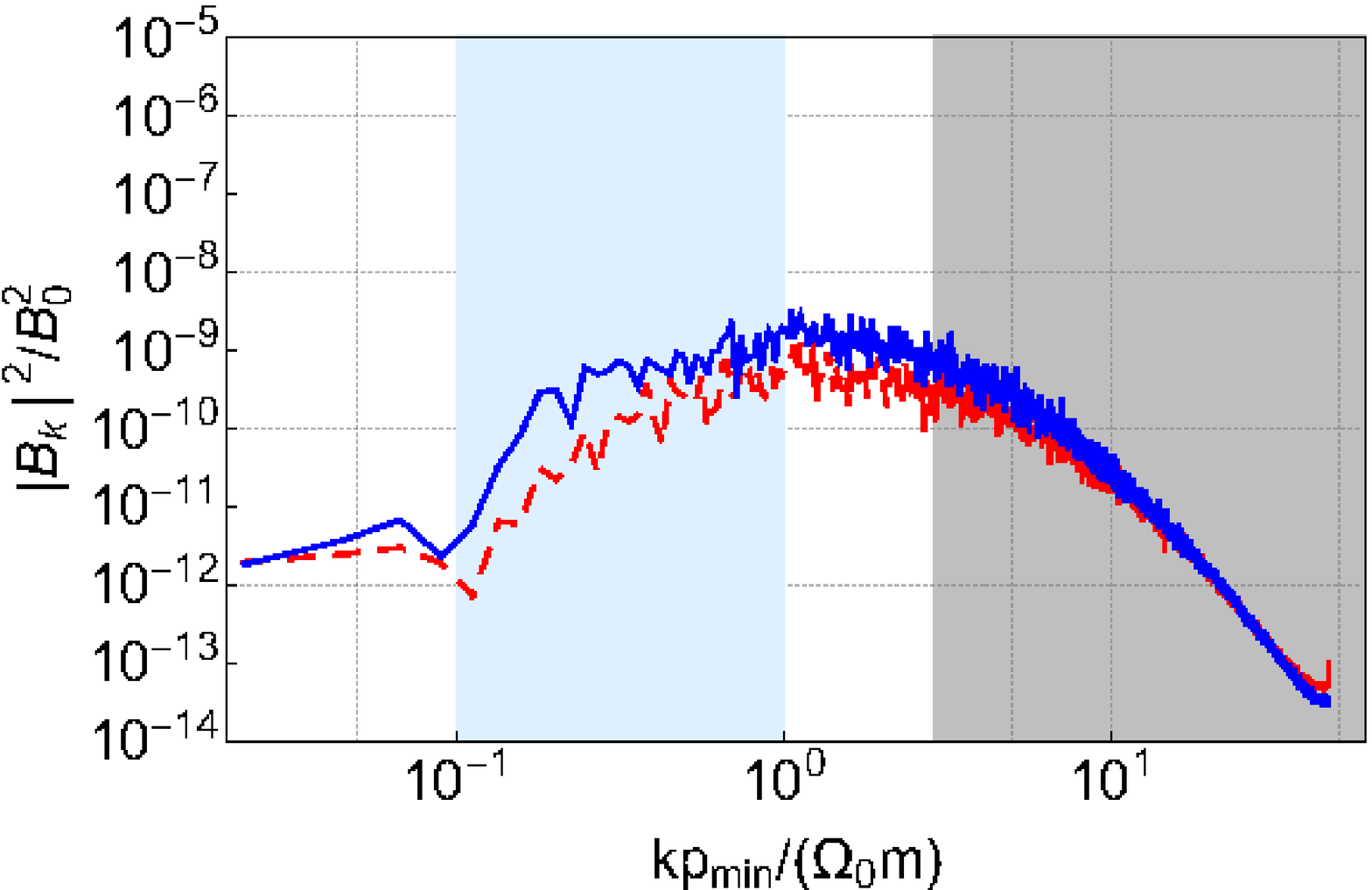} &
\includegraphics[width=\columnwidth]{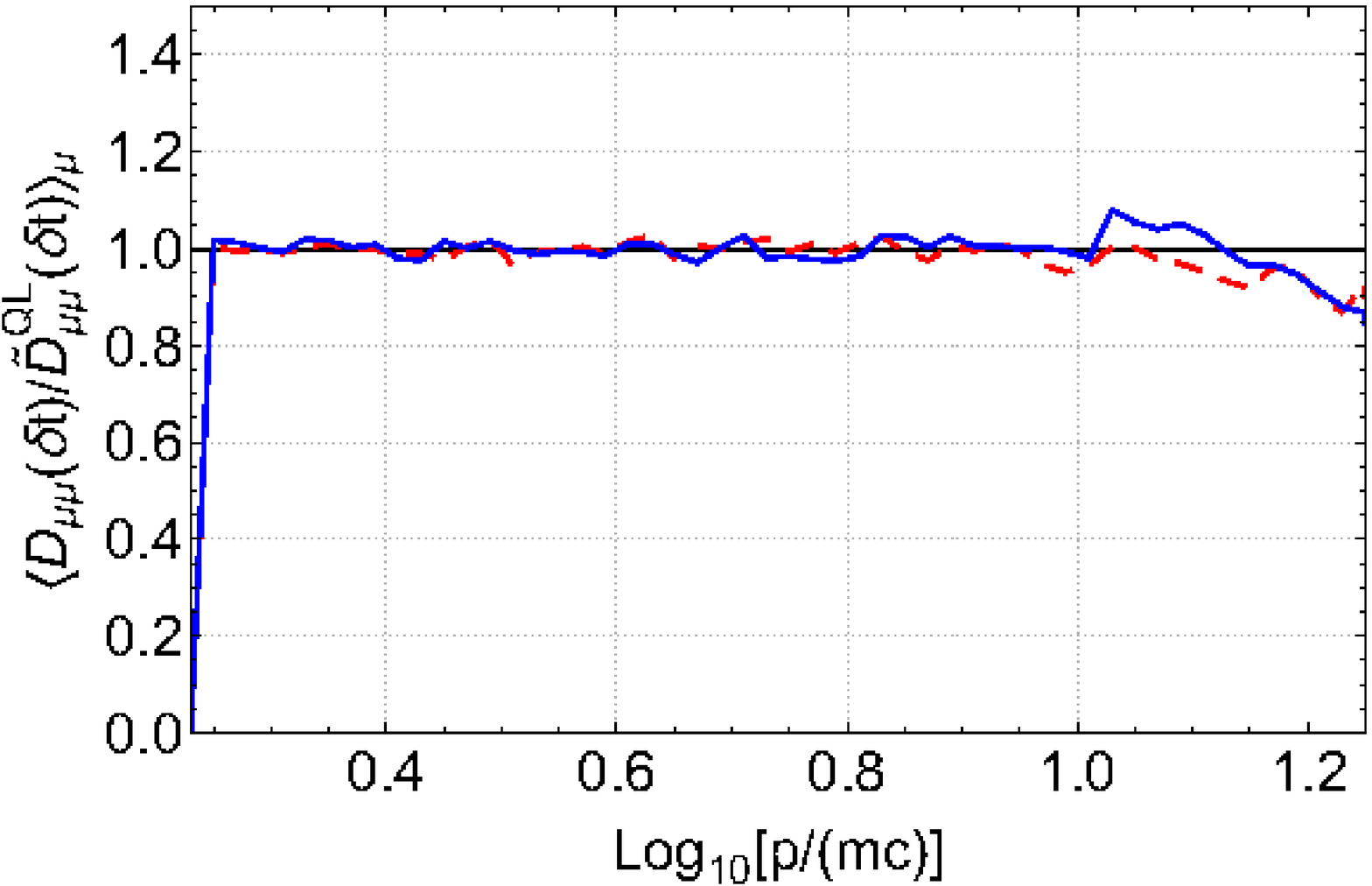} \\
\includegraphics[width=\columnwidth]{./figs/l2-skip99-BK.eps} &
\includegraphics[width=\columnwidth]{./figs/l2-skip99-PPitchangle.eps} \\
\includegraphics[width=\columnwidth]{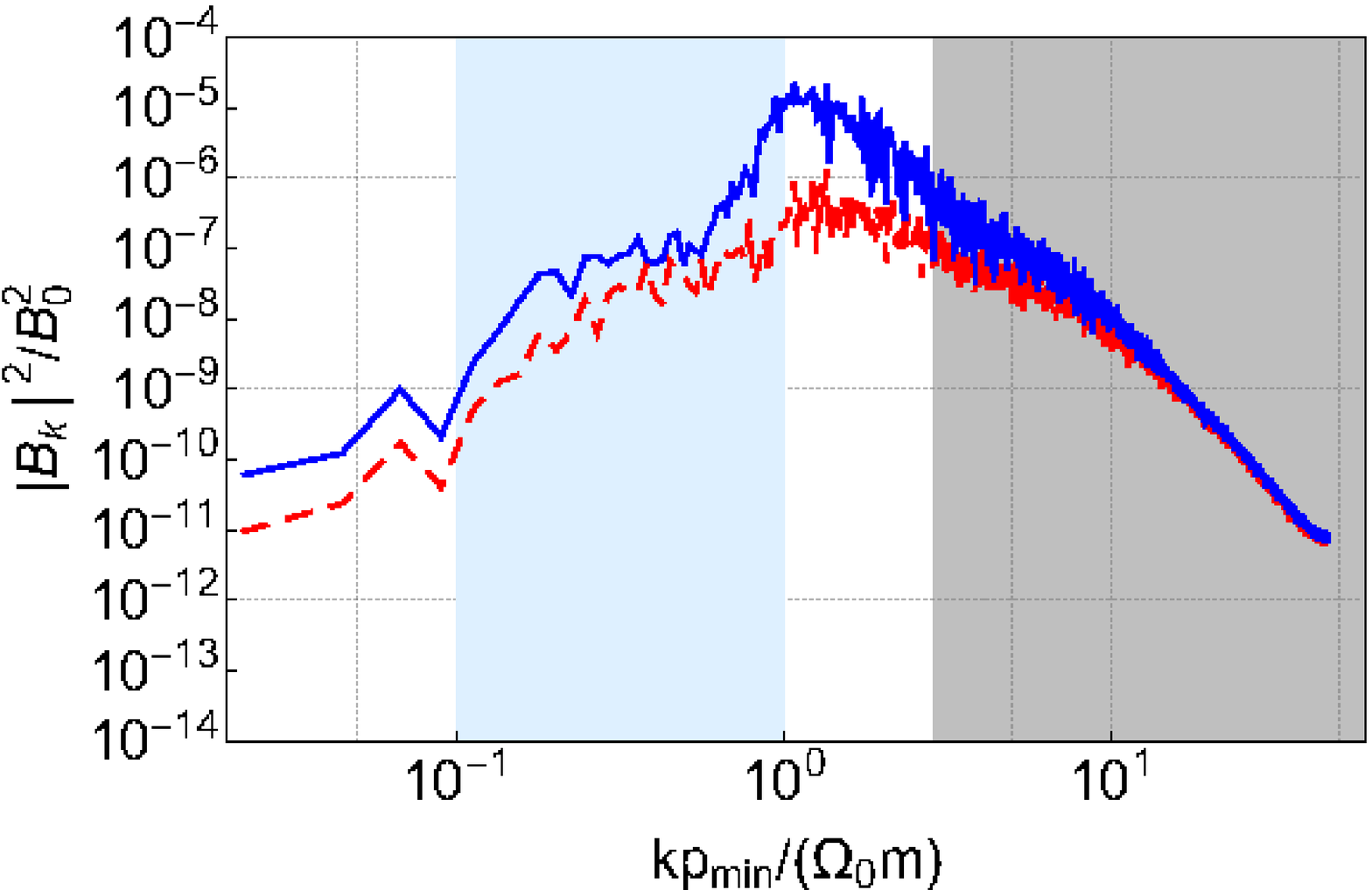} &
\includegraphics[width=\columnwidth]{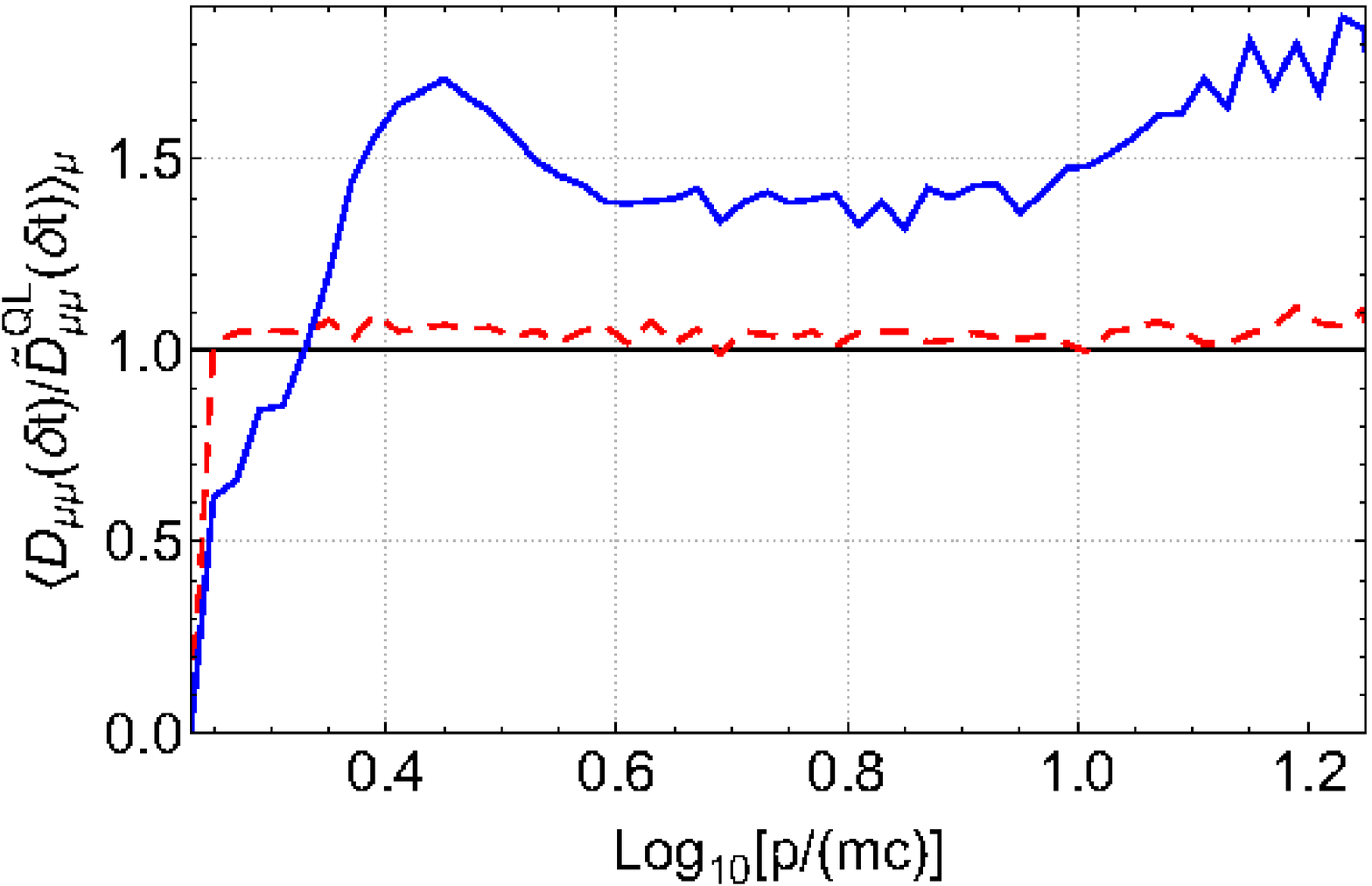}
\end{tabular}
\caption{Same as Fig.~\ref{fig:short_anis0}, for models 
with different $n_{cr}/n_i:$ $2.5 \times 10^{-5}$ ({\it top row}), $10^{-4}$ ({\it middle row}), 
$4 \times 10^{-4}$ ({\it lower row}), which correspond respectively to models $l4$, $l2$, and $l5$ of Table~\ref{tab:runs_param}.}
\label{fig:short_ncroni}
\end{figure*}

Figure~\ref{fig:short_ncroni} is similar to Figure~\ref{fig:short_anis0}, but it shows a comparison between models with fixed anisotropy $A_0$ and different values 
of $n_{cr}/n_i$ (models $l2$, $l4$, and $l5$ in Table~\ref{tab:runs_param}). 
While the models with lower growth rate of the instability $|\Gamma_{\max}|$ (models $l2$ and $l4$) present the 
better agreement between $D_{\mu\mu}(\delta t)$ and $\widetilde{D}^{QL}_{\mu\mu} (\delta t)$, 
the model with higher $n_{cr}/n_i$ (model $l5$) shows a diffusion rate $D_{\mu\mu}(\delta t)$ 
about two times larger than $\widetilde{D}^{QL}_{\mu\mu} (\delta t)$
for the final time.

\subsection{Instability saturation}

\begin{figure}
\centering 
\includegraphics[width=\columnwidth]{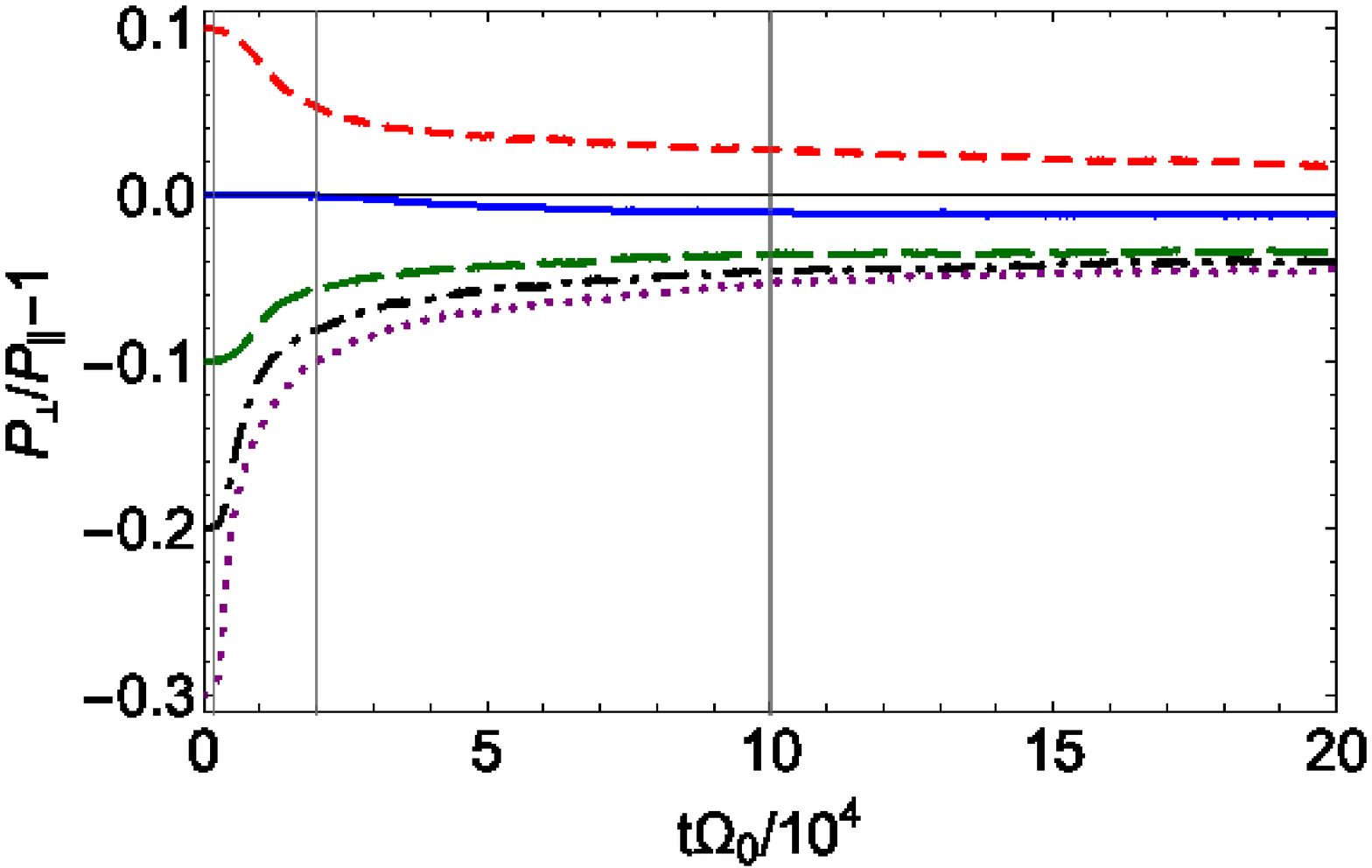} \\
\includegraphics[width=\columnwidth]{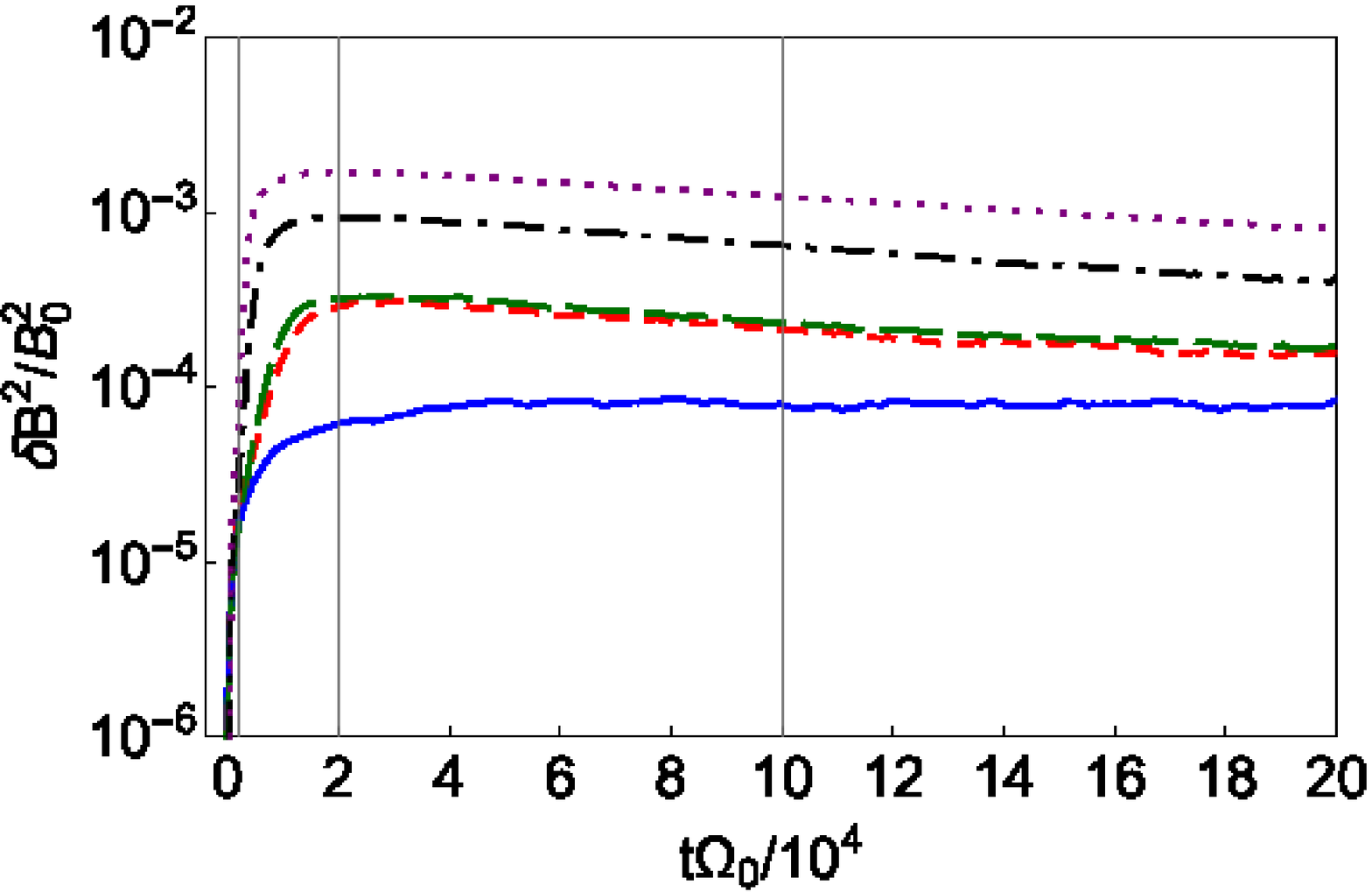}
\caption{Time evolution of the total magnetic energy in fluctuations normalized by the 
energy in the mean field (upper panel) and CR total 
pressure anisotropy (bottom panel) for models starting with different initial anisotropies 
$A_0:$ $+0.1$ ({\color{red} red dashed line}), $0$ ({\color{blue} blue solid line}), $-0.1$ ({\color{darkgreen} green long-dashed line}), $-0.2$ (black dot-dashed line), $-0.3$ ({\color{magenta} magenta dotted line}), 
which correspond to models $s1$-$s5$ in Table~\ref{tab:runs_param}. 
The vertical lines mark four different stages of the instability evolution: linear, early saturation, 
and late saturation times (from smaller to larger $t$).}
\label{fig:long_evol}
\end{figure}

\begin{figure}
\includegraphics[width=\columnwidth]{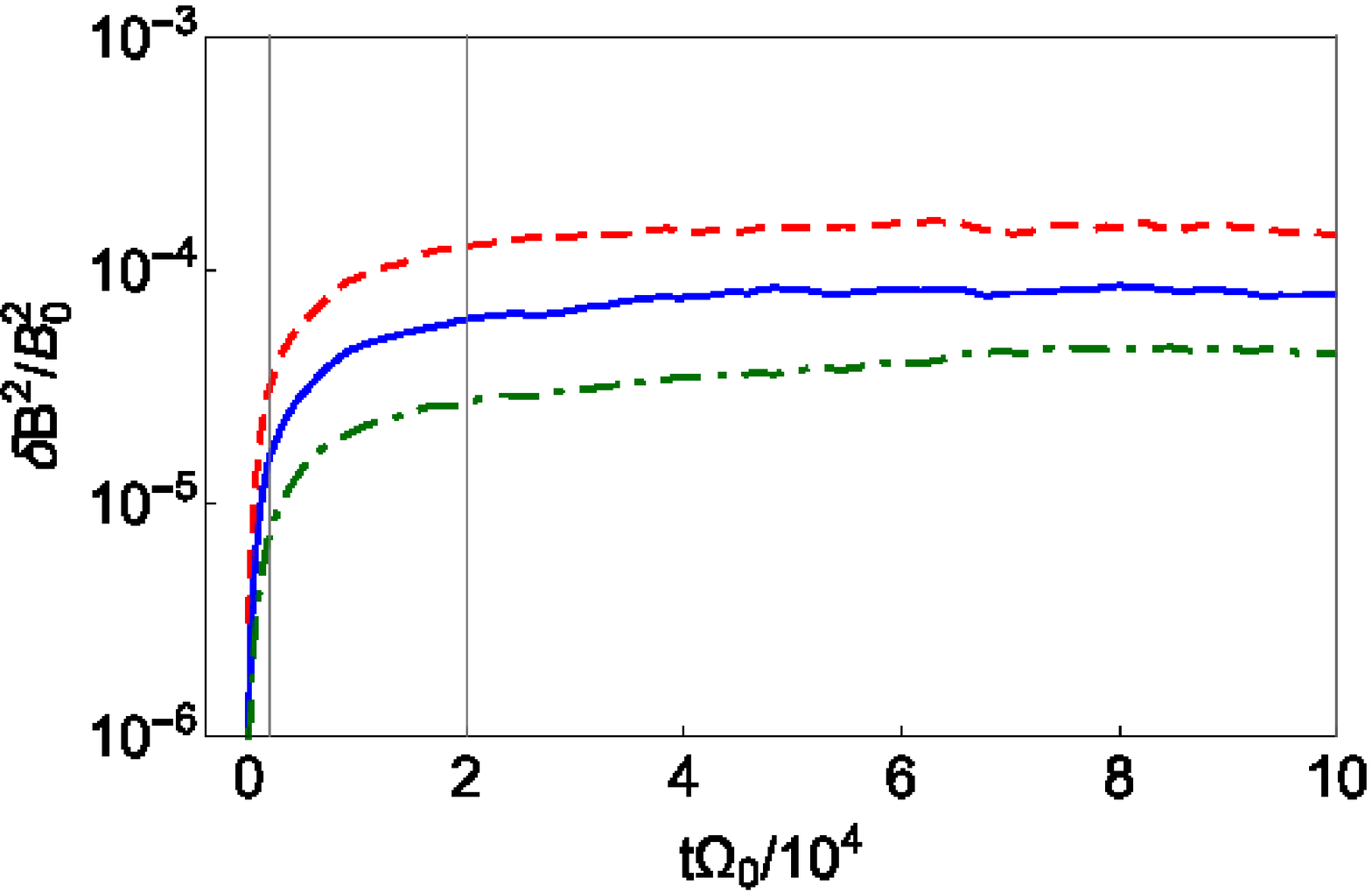}
\caption{Time evolution of the total magnetic energy in fluctuations normalized by the 
energy in the mean field 
for models starting with initial anisotropies 
$A_0 = 0$ and different number of particles NP/NX: 512 ({\color{red} red dashed line}), 1024 ({\color{blue} blue solid line}), 2048 ({\color{darkgreen} green dot-dashed line}), 
which correspond respectively to models $s6$, $s2$, and $s7$ in Table~\ref{tab:runs_param}.}
\label{fig:long_evol_np}
\end{figure}

Figure~\ref{fig:long_evol} shows the time evolution of the CR pressure anisotropy and of the total magnetic energy 
of the waves (normalized by the energy of the mean field) 
for models $s1$-$s5$
starting with different anisotropies $A_0$ (see Table~\ref{tab:runs_param}). It shows clearly the reduction of the absolute 
total anisotropy, although the values do not achieve zero by the end of our simulations. 
One control simulation with initial zero anisotropy (model $s2$) is shown to be stable and to continue isotropic
until the end of the simulation. 
After $t \Omega_0 \sim 10^4$, 
the total magnetic energy in the fluctuations begins to saturate, 
at higher values for the models with higher $|A_0|$. The vertical lines in Figure~\ref{fig:long_evol} 
indicate the times in different regimes of the instability:
$t \Omega_0 = 2 \times 10^{3}$ ({\it linear phase}),
$t \Omega_0 = 2 \times 10^{4}$ ({\it beginning of the saturation phase}), and
$t \Omega_0 = 10^{5}$ and
$t \Omega_0 = 2 \times 10^{5}$ ({\it late times during the saturation phase}).
We observe that the energy of the waves in the isotropic case also increases initially and saturates, 
caused by the numerical noise due to the limited number of particles. This value is smaller for higher number of particles (see
Figure~\ref{fig:long_evol_np} for a comparison between simulations with different NP/NX).
It indicates the minimum limit in the instability growth rate that we can simulate (with fixed resolution), below which 
the properties are dominated by the numerical noise.


\begin{figure*}
\begin{tabular}{c c} 
\includegraphics[width=\columnwidth]{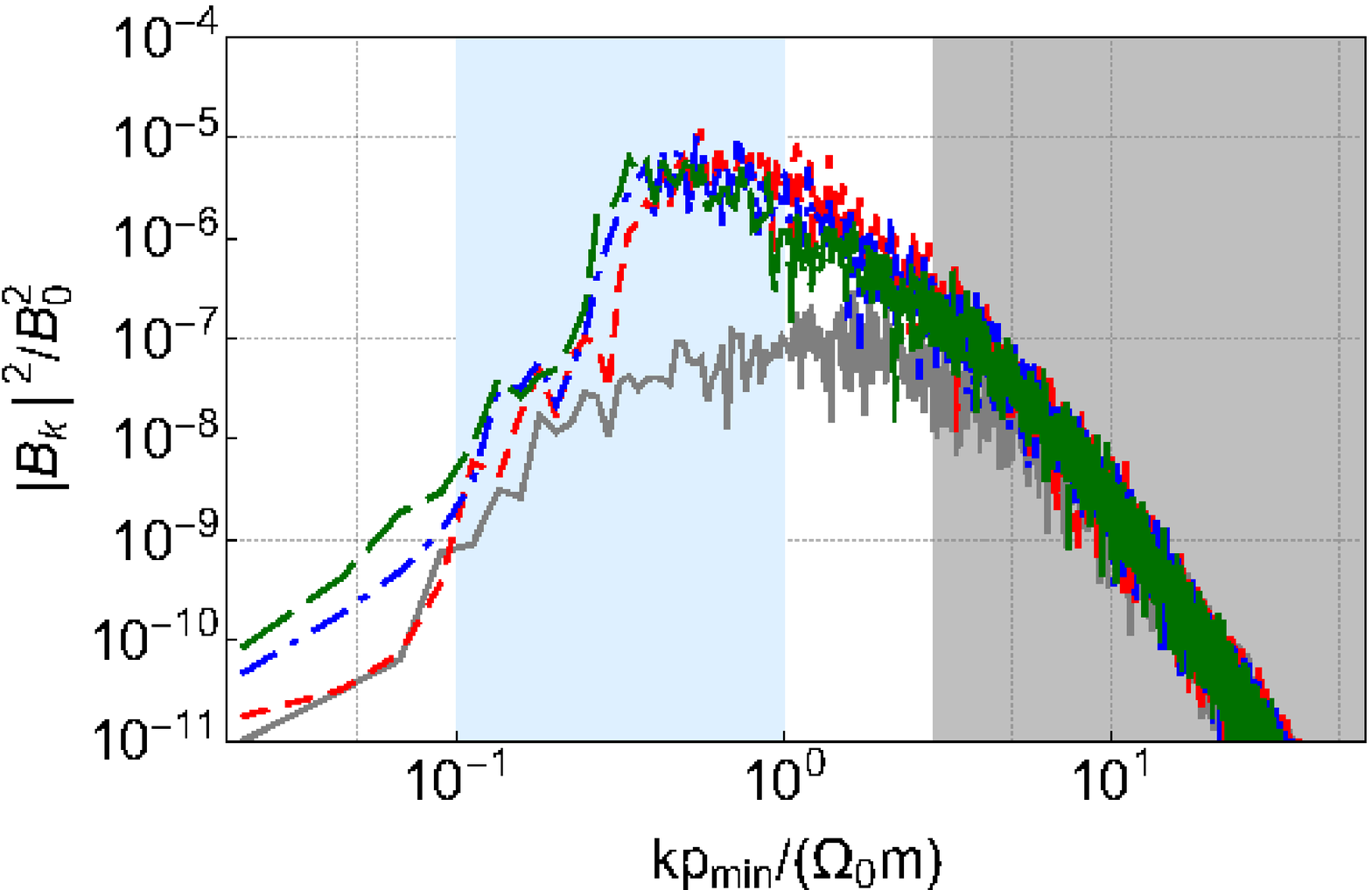} &
\includegraphics[width=\columnwidth]{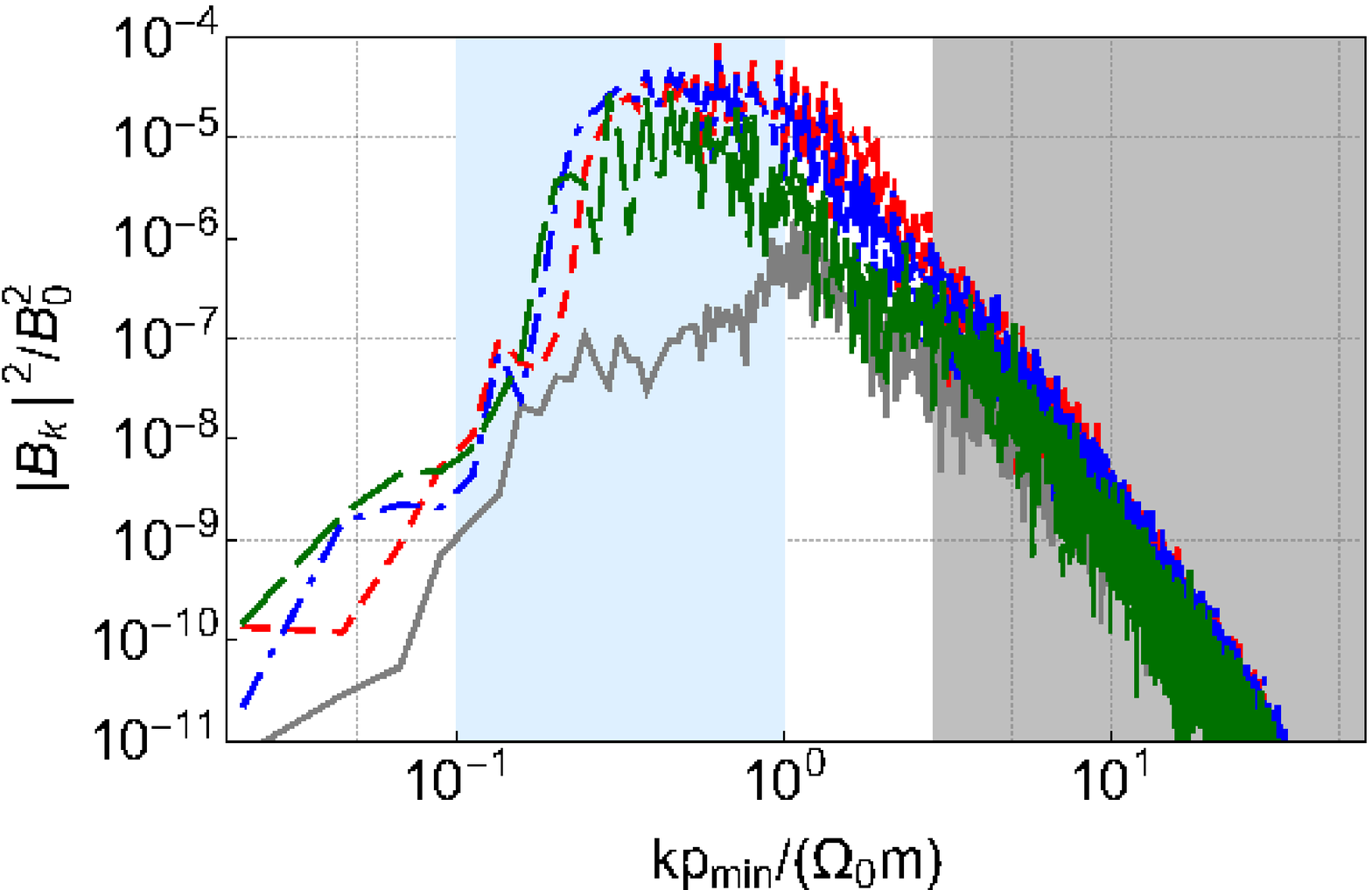} \\
\includegraphics[width=\columnwidth]{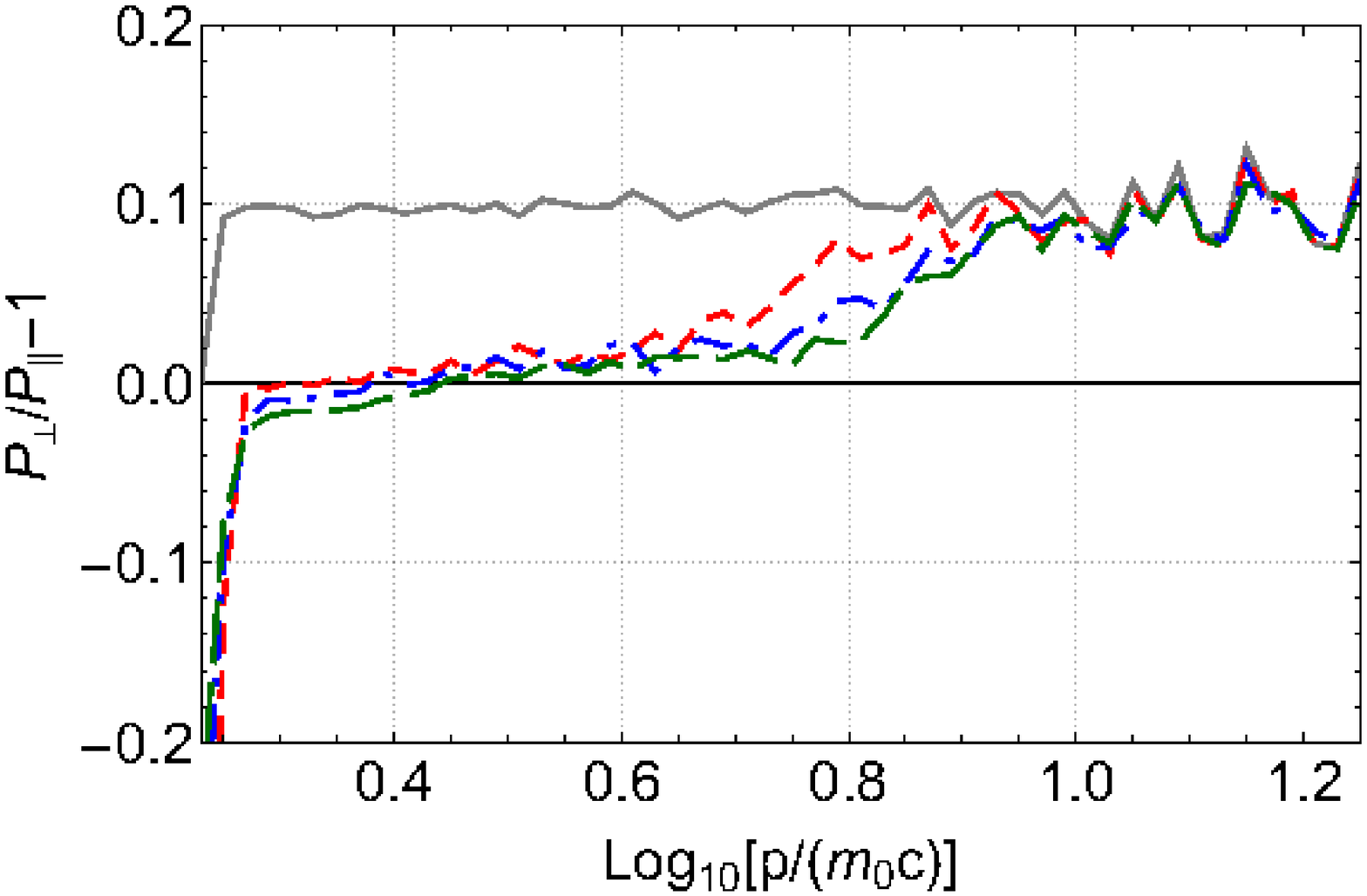} &
\includegraphics[width=\columnwidth]{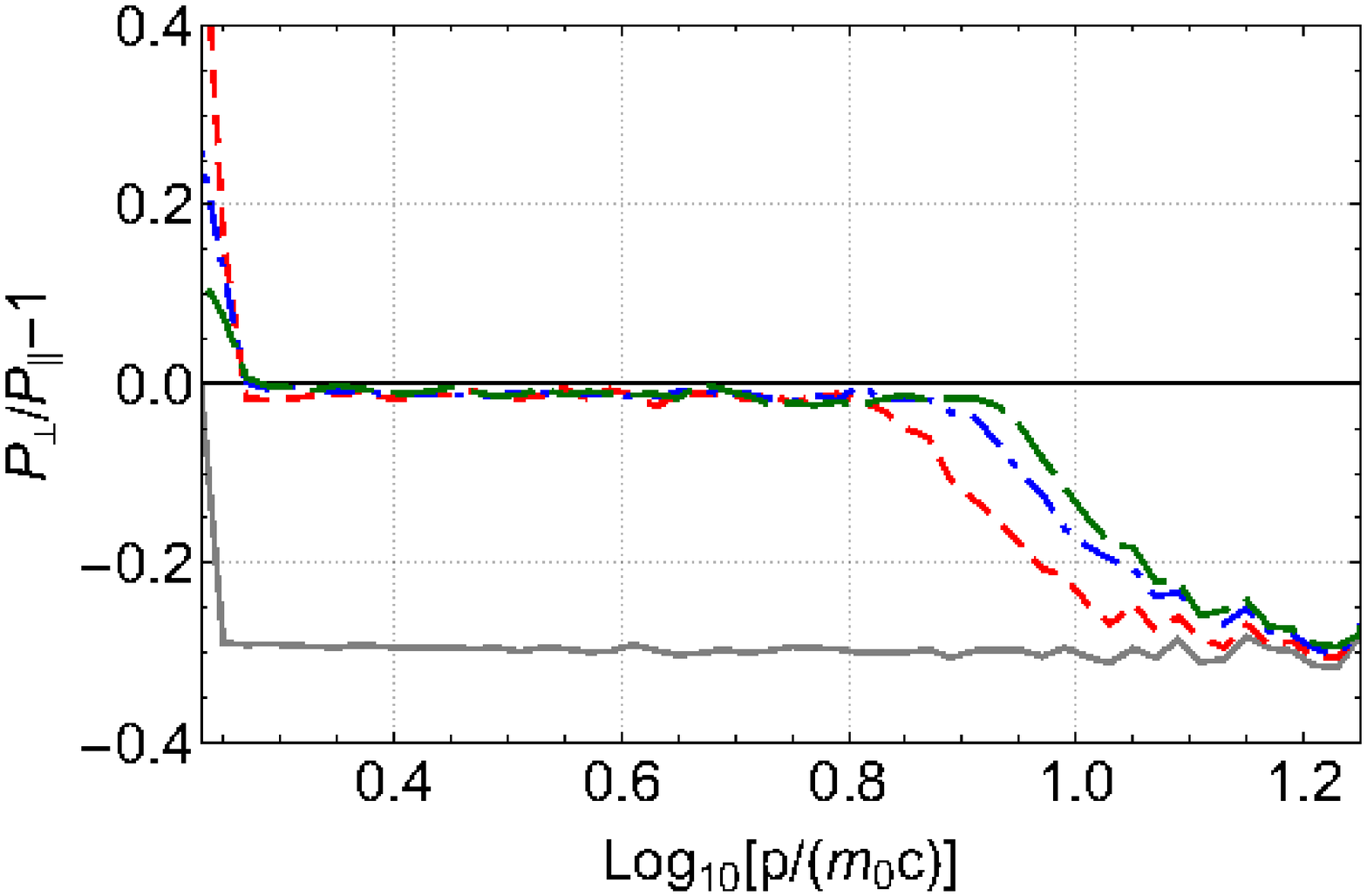} \\
\includegraphics[width=\columnwidth]{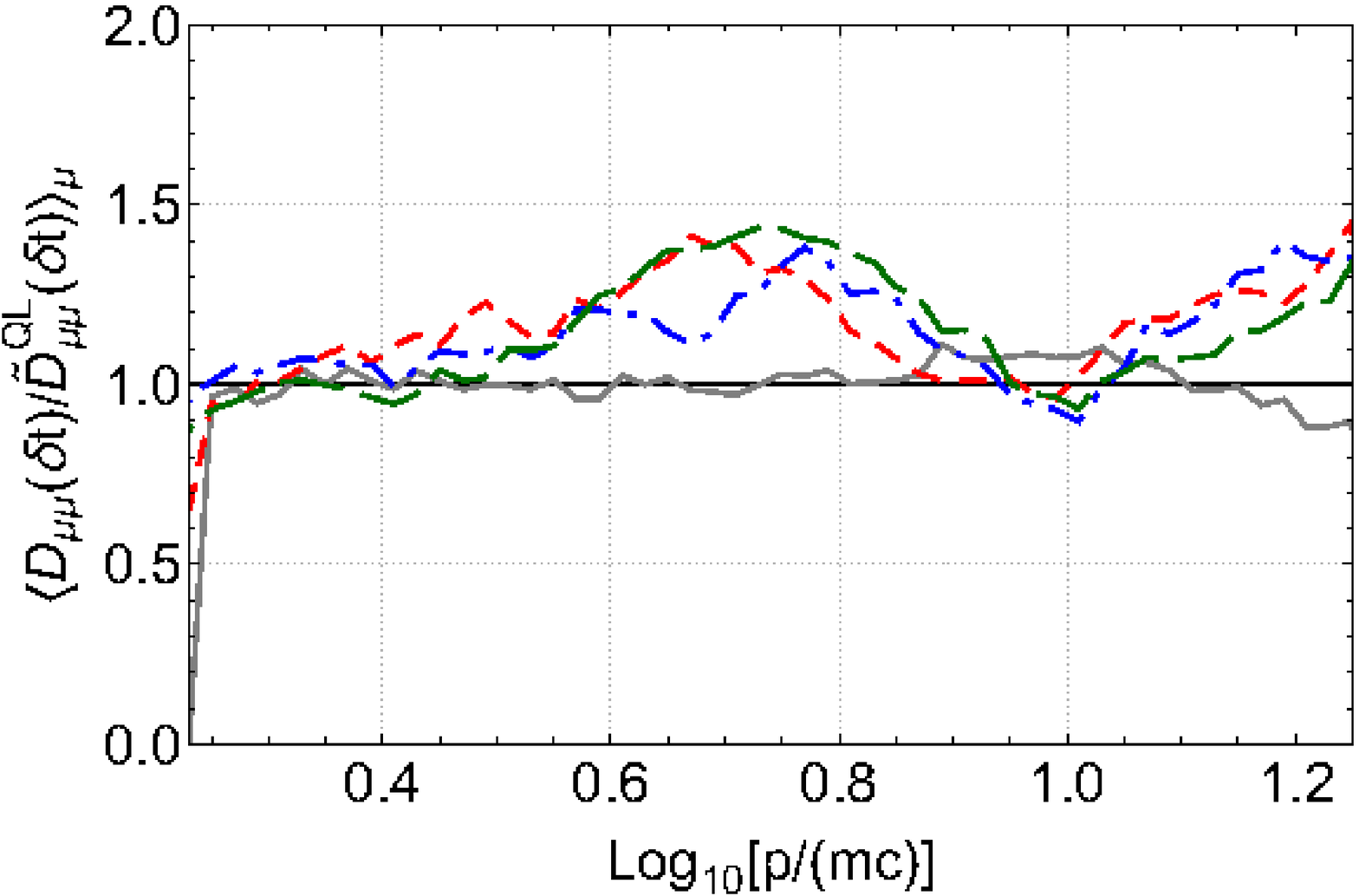} &
\includegraphics[width=\columnwidth]{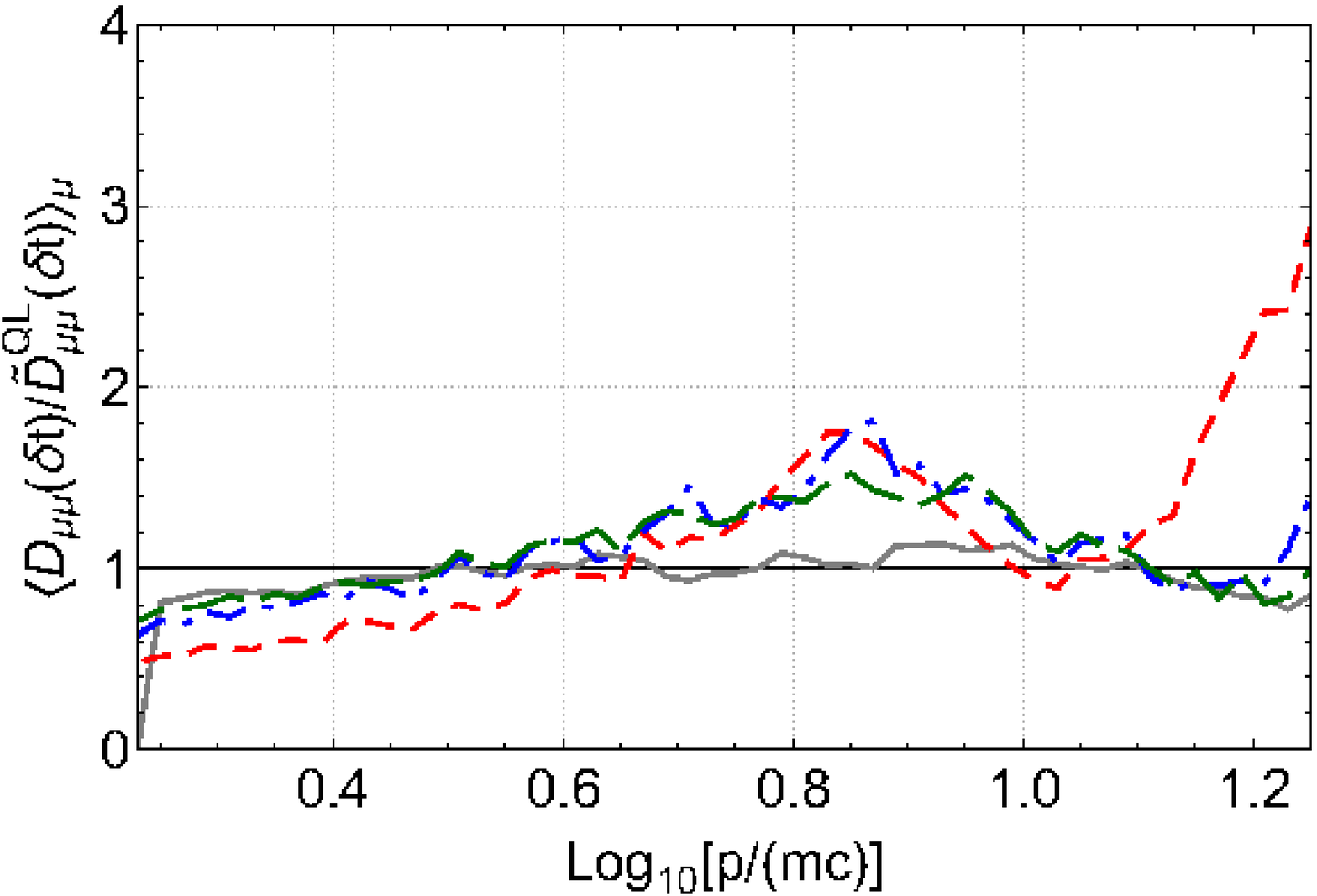}
\end{tabular}
\caption{Normalized power spectrum of magnetic field $|B(k)|^2/B_0^2$ (upper row),  
$\langle D_{\mu\mu}(\delta t) / \widetilde{D}_{\mu\mu}^{QL} (\delta t) \rangle_{\mu}$ ratio
(middle row), and distribution of CR pressure anisotropy $A$ in momentum (bottom row)
for models $s1$ and $s5$ (see Table~\ref{tab:runs_param})
with different anisotropies $A_0$: $+0.1$ (left column) and $-0.3$ (right column).
The lines with different colors 
indicate different times (shown in Figure~\ref{fig:long_evol}): 
$t \Omega_0 = 2 \times 10^3$ (gray solid line), 
$t \Omega_0 = 2 \times 10^4$ ({\color{red} red dashed line}), 
$t \Omega_0 = 10^5$ ({\color{blue} blue dot-dashed line}),
$t \Omega_0 = 2 \times 10^5$ ({\color{darkgreen} green long-dashed line}). In the power spectrum plots (upper row), 
the gray area indicates the wave number interval
where numerical dissipation dominates 
(wavelengths $\le 32$ grid cells); the blue area represents the wave number interval $\Omega_0 m_{cr} / p_{\max} < k < \Omega_0 m_{cr} / p_{\min}$.} 
\label{fig:long_spectrum}
\end{figure*}

We show in Figure~\ref{fig:long_spectrum},
the magnetic power spectrum, the ratio 
$\langle D_{\mu\mu}(\delta t)/\widetilde{D}^{QL}_{\mu\mu}(\delta t) \rangle_{\mu}$,
and the anisotropy distribution in the momentum $A(p)$, for two different models ($A_0 = +0.1$ and $A_0 = -0.3$). 
We plot each quantity at the four times (indicated by different colors) of the different regimes 
described above (see also Figure~\ref{fig:long_evol}). 
In the beginning of the saturation phase, the magnetic energy 
in the blue-shaded region is already saturated for the highest wave-numbers, but it is still growing for the smallest 
wave numbers.

The diffusion coefficients in Figure~\ref{fig:long_spectrum} are calculated using time intervals $\delta t \Omega_0 = 10^3$
($\delta t \Gamma_{\max} \approx 1.4$ for model $s5$ which has the highest $\Gamma_{\max}$).
During the linear phase, $D_{\mu\mu}(\delta t)$ compares well with $\widetilde{D}^{QL}_{\mu\mu} (\delta t)$.
For the later times these values
are still comparable by a factor of 3.
The fluctuation is stronger for the higher initial anisotropy model $A_0 = +0.3$, at the beginning of the saturation phase 
($t \Omega_0 = 2 \times 10^4$, red dashed line in Figure~\ref{fig:long_spectrum}), when the magnetic energy spectrum is more irregular 
inside the wavenumber interval of main resonance (blue area).
This effect can be caused by the shortness of the time interval $\delta t$ compared to the correlation 
time $t_c$ of the rate of change of the pitch-angle (see Appendix~\ref{sec:appendix}) for the particles of higher energies, 
as the deviation is larger for these particles.

\begin{figure}
\begin{tabular}{c} 
\input{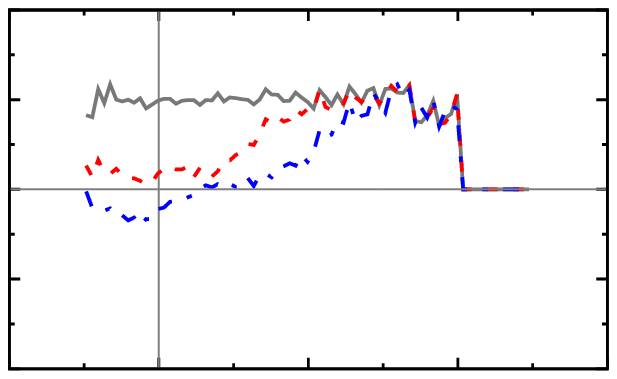} \\
\input{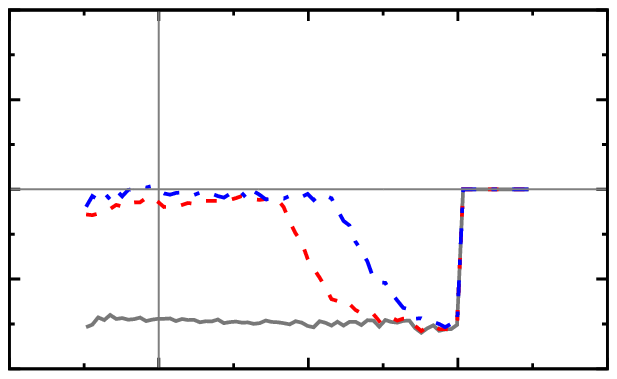}
\end{tabular}
\caption{Distribution of the pressure anisotropy 
in momentum $A(p)$ for the modified models $s1$ (upper panel) 
and $s5$ (bottom panel).
The modification consists in the extension of the CRs distribution, originally given by a single power-law ($\propto p^{-2.8}$) 
in the compact momentum interval $[p_{\min}, p_{\max}]$, to include the low momentum interval $[0.1 p_{\min}, p_{\min}]$ 
where it assumes a steep, growing power-law ($\propto p^{5.6}$).
The lines with different colors 
indicate different times: 
$t \Omega_0 = 2 \times 10^3$ (gray solid line), 
$t \Omega_0 = 2 \times 10^4$ ({\color{red} red dashed line}), 
$t \Omega_0 = 10^5$ ({\color{blue} blue dot-dashed line}). }
\label{fig:long_modified}
\end{figure}

For the initial linear phase the anisotropy distribution of both models is shown to be almost identical to the initial one.
In the beginning of the saturation phase, however, the low-energy CRs (the bulk of the CRs distribution) are already 
almost totally isotropized (at the same time the instability growth drops for the corresponding resonant wavenumbers). 
The relatively strong fluctuations in $A$ seen at $p \approx p_{\min}$ is an artifact caused by the 
discontinuity at $p=p_{\min}$ of the distribution function $f(p)$ employed (Eq.~\ref{eq:distr_p}). These peaks disappear when we repeat the simulations extending 
$f(p)$ towards smaller $p$ with a steep, growing power law (Figure~\ref{fig:long_modified}). The single power law in the interval $[p_{\min}, p_{\max}]$, 
however, has the advantage of the simplicity in the analytical treatment.
At the final time of the simulation, 
the CRs of the highest energies still preserve the initial anisotropy.

\begin{figure}
\includegraphics[width=\columnwidth]{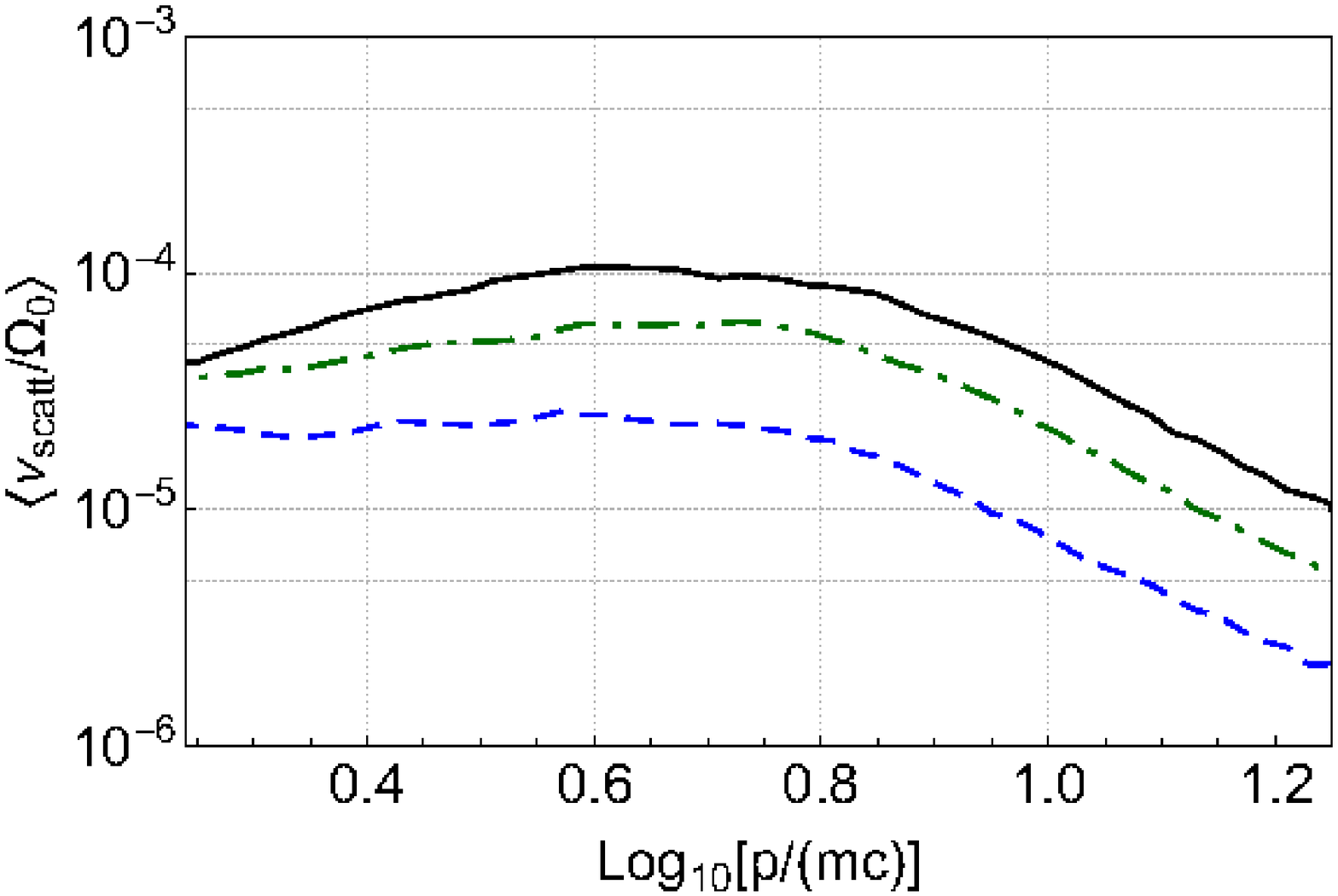}
\caption{Distribution of the scattering rate $\nu_{scatt} = \left\langle 2 D_{\mu\mu}(\delta t)/(1 - \mu^2) \right\rangle_{\mu}$ (normalized by $\Omega_0$) 
at time $t \Omega_0 = 2 \times 10^5$ (later time during the saturation phase, see last vertical line in Figure~\ref{fig:long_evol}) for models with different initial anisotropy $A_0 = -0.1$ ({\color{blue} blue dotted line}), $-0.2$ ({\color{darkgreen} green dot-dashed line}), $-0.3$ (black solid line), corresponding to models $s3$-$s5$ in Table~\ref{tab:runs_param}.}
\label{fig:long_scatt}
\end{figure}

In Figure~\ref{fig:long_scatt} we show 
the scattering rate 
$\nu_{scatt} \equiv \langle 2D_{\mu\mu}(\delta t)/(1 - \mu^2) \rangle_{\mu}$
during the saturated phase 
(at the final time of the simulation)
for three models $A_0 = -0.1, -0.2, -0.3$. 
The distribution as a function of the CR momentum is almost flat for the 
momentum range which is isotropized by this time, but decaying for larger energies. 
This is expected as the high-energy 
particles are resonant with waves which are still growing.

\section{Summary and Conclusions}\label{sec:summary}

Using one-dimensional hybrid PIC-MHD simulations, we study numerically the evolution of the 
CR gyroresonance instability, triggered by a distribution
of CR protons with initial anisotropy (with respect to the local mean magnetic field direction)
in pressure ($P_{cr,\parallel} \ne P_{cr,\perp}$). We restricted our analysis to parallel-propagating 
modes, 
which are the 
fastest growing modes. 
During the linear phase of the instability 
the growth rate and phase speed of the modes with 
Right and Left 
circular polarization 
show excellent agreement with the theoretical dispersion relation, 
for both initial setups with
$P_{cr,\parallel}>P_{cr,\perp}$ and $P_{cr,\parallel}<P_{cr,\perp}$.

In all our simulations the non-linear wave-particle effects are important. After a short initial 
period of exponential growth of the waves, the scattering and consequent isotropization of the CRs 
momentum distribution is the mechanism which gradually saturates the instability growth. 
The low-energy CRs are isotropized faster than 
those of the higher energies.
The amplitude of the waves and the scattering rate of particles 
during the saturation phase are larger for initially larger maximum instability growth rate $\Gamma_{\max}$.

We extracted from the simulations the pitch angle diffusion coefficient $D_{\mu\mu}$ for the evolution of the CR distribution 
function averaged over a time-scale $\delta t \lesssim \Gamma_{\max}^{-1}$, and we find the empirical values in good 
agreement with the QLT estimates for static waves (within a factor of 3).
This agreement is shown to be better for our simulations with smaller $\Gamma_{\max}$.
Indeed, due to limitations 
imposed by the noise caused by the low sample of macro-particles in the PIC technique,
all our simulations have parameters which produce a maximum growth rate of the instability much higher than 
expected in realistic situations.
Nonetheless, this direct confirmation of the applicability 
of the QLT to estimate the CR scattering by the gyroresonance instability is a valuable 
support for theoretical models 
connecting the large-scale turbulence cascade with the ``microphysics'' of the CR instabilities
(\citetalias{yan_lazarian_2011}),
and for subgrid models  
in large-scale simulations involving CR transport (for example \citealt{everett_zweibel_2011, evoli_yan_2014, delvalle_etal_2015, pfrommer_etal_2017}),
as direct numerical simulations cannot cover the huge range of scales involved.

In conclusion, the outcome of this work provide a solid foundation for developing further investigations on the role 
of the CR gyroresonance instability on CR propagation.
This includes for example the use of a setup where the pressure anisotropy is generated naturally by continuous 
compression or shear (e.g. \citealt{kunz_etal_2014, riquelme_etal_2015, sironi_narayan_2015}), 
the combined effect with the streaming instability generated by the presence of CR drift,
and the use 
of three-dimensional simulations which would better represent the turbulence cascade from the instability 
generated waves, necessary for understanding the decaying of the instability and 
the effect of successive large-scale random compressions/expansions provided by the large-scale turbulence 
(\citealt{melville_etal_2016}).
As already stressed before, 
the precise quantitative knowledge of the scattering provided by microscopic instabilities is 
an indispensable ingredient for building more realistic models 
for the CR propagation in the Galaxy and in the ICM, needed for our correct understanding and interpretation
of almost all high-energy phenomena.

\section*{Acknowledgements}
\addcontentsline{toc}{section}{Acknowledgements}

O.L. acknowledges the DESY Summer Student Programme 2016, during which part of this project was developed.
R.S.L. acknowledges A. Beresnyak, E.M. de Gouveia Dal Pino, C.-S. Jao, Y. Mardoukhi, and M. Pohl for 
very useful comments and discussions; he is also indebted to M.V. del Valle and S. Vafin for their invaluable detailed suggestions.
The authors would like to thank the anonymous referee who helped to improve this work with his/her comments and suggestions.







\clearpage

\appendix

\section{Quasilinear Fokker-Planck pitch-angle diffusion coefficient}
\label{sec:appendix}

Consider the operational definition of the pitch-angle diffusion coefficient convenient to analyse the particles 
simulation:
\begin{equation}
D_{\mu\mu} (\mathbf{X}(t_0), t_0) = \frac{\left\langle \left[ \mu(t) - \mu(t_0) \right]^2 \right\rangle}{2 (t - t_0)},
\end{equation}
where the angle brackets $\langle \cdot \rangle$ represent an ensemble over 
all the particles with coordinates $\mathbf{X} = (X, Y, Z, \mu, p, \psi)$ at time $t_0$,
where $(X, Y, Z)$ 
are the coordinates of the guiding center, and $(p, \mu, \psi)$ are the momentum, cosine of pitch-angle, and gyro-phase, respectively.
In what follows, we provide the derivation of $D_{\mu\mu}$ 
for particles travelling in an homogeneous plasma superimposed by a spectrum of 
small amplitude transverse electromagnetic waves, with circular polarization and
propagating parallel to the mean magnetic field.
We also show that the above definition coincides --- using similar hypotheses --- with the 
$D_{\mu\mu}$ coefficient in the Fokker-Planck equation for the evolution of the (ensemble) averaged distribution function 
derived  through the quasilinear theory (see e.g. \citealt{achatz_etal_1991, schlickeiser_2002}).

Without loss of generality, we take $t_0 = 0$ from here:
\begin{equation}
\begin{split}
D_{\mu\mu} &= \frac{\left\langle \left[ \mu(t) - \mu(0) \right]^2 \right\rangle}{2 t} \\
&= \frac{1}{2 t} \left\langle \int_{0}^{t} ds \dot{\mu}'(s) \int_{0}^{t} d\xi \dot{\mu}'(\xi) \right\rangle \\
&= \frac{1}{2 t} \int_{0}^{t} ds \int_{-s}^{t - s} d\tau \left\langle \dot{\mu}'(s) \dot{\mu}'(s + \tau) \right\rangle, \\
\end{split}
\label{eq:dmumu_1}
\end{equation}
where the prime means the integration is to be performed along the unperturbed particles orbit coordinates 
$\mathbf{X}'$.
Adopting a cartesian coordinate frame with the $x-$axis pointing 
in the direction of the mean magnetic field $\mathbf{B_0}$, 
\begin{equation}
\begin{split}
X' &= X + v \mu s, \\
Y' &= Y, \\
Z' &= Z, \\
p' &= p, \\
\mu' &= \mu, \\
\psi' &= \psi - \varepsilon \Omega s,
\end{split}
\label{eq:orbit}
\end{equation}
where $v = p/\gamma m$ is the particle speed, $\gamma = \left(1 - v^2/c^2 \right)^{-1/2}$, $c$ is the light speed, 
$m$ is the rest mass of the particles, $\varepsilon = q/|q|$ is the signal of the particles charge, and $\Omega = |q| B_0/\gamma m c$
is the particles cyclotron frequency (the unprimed phase-space coordinates are to be understood at time $t = 0$).

Assuming the existence of a correlation time $t_c$ so that the 
correlation $\left\langle \dot{\mu}'(s) \dot{\mu}'(s + \tau) \right\rangle$
decays fast for $|\tau| > t_c$,
then the main contribution to the integral~\ref{eq:dmumu_1} comes from the interval $-t_c \le \tau \le t_c$. 
For $t \gg t_c$, the intervals of the second integral can be replaced by $-t$ to $t$:
\begin{equation}
\begin{split}
D_{\mu\mu} &\approx \frac{1}{2 t} \int_{0}^{t} ds \int_{-t}^{t} d\tau \left\langle \dot{\mu}'(s) \dot{\mu}'(s + \tau) \right\rangle \;\;\; (t \gg t_c)\\
&\approx \frac{1}{t} \int_{0}^{t} ds \int_{0}^{t} d\tau \left\langle \dot{\mu}'(s) \dot{\mu}'(s + \tau) \right\rangle, \\
\end{split}
\label{eq:dmumu_2}
\end{equation}
where in the last approximation we assumed the dependence of the correlation in $\tau$ to be approximately symmetric around zero.

Additionally, if the correlation $\left\langle \dot{\mu}'(s) \dot{\mu}'(s + \tau) \right\rangle$ does not depend on $s$, or if it changes 
on a time scale $t^* \gg t$,
\begin{equation}
D_{\mu\mu} \approx \int_{0}^{t} d\tau \left\langle \dot{\mu}'(0) \dot{\mu}'(\tau) \right\rangle.
\end{equation}
which coincides with the expression obtained from the quasilinear theory.
In the remaining of this Appendix, we work out one expression for $D_{\mu\mu}$ in terms of the 
waves spectrum description. 

Writing the transverse magnetic fluctuations in wave components with Left ($L$) or Right ($R$) circular polarization:
{\footnotesize
\begin{equation}
\begin{split}
\delta B_y (x, t)
&= \sum_{n} \int_{-\infty}^{\infty} dk \delta B_{y, n} (k, x, t) \\
&= \sum_{n} \int_{-\infty}^{\infty} dk \frac{1}{2} \left[ \delta A_{n}^L (k, x, t) + \delta A_{n}^R (k, x, t) \right]
\end{split}
\label{eq:by}
\end{equation}
\begin{equation}
\begin{split}
\delta B_z (x, t)
&= \sum_{n} \int_{-\infty}^{\infty} dk \delta B_{z, n} (k, x, t) \\
&= \sum_{n} \int_{-\infty}^{\infty} dk \frac{i}{2} \sign (l_n k) \left[ - \delta A_{n}^L (k, x, t) + \delta A_{n}^R (k, x, t) \right]
\end{split}
\label{eq:bz}
\end{equation}
}
where $l_n$ is the direction of propagation of the wave, $n=b,f$ are waves propagating backward ($n=b$, $l_n=-1$) and forward ($n=f$, $l_n=1$) 
to the mean magnetic field direction. The components $\delta A_n^{\alpha}$ ($\alpha=L,R$) are monocromatic waves given by
\begin{equation}
\begin{split}
\delta A_{n}^{\alpha}(k, x, t) 
&= A_{n}^{\alpha} (k, t) \exp \left\{ i \left[ kx - \omega_{r,n}^{\alpha}(k) t \right] \right\} \\
&= | A_{n}^{\alpha} (k, t) | \exp \left[ i \phi_{n}^{\alpha} (k) \right] \exp \left\{ i \left[ kx - \omega_{r,n}^{\alpha} (k) t \right] \right\}, \\
\end{split}
\end{equation}
with
\begin{equation}
| A_{n}^{\alpha} (k, t) | = | A_{n}^{\alpha} (k, 0) | \exp \left[ \Gamma_n^{\alpha}(k) t \right]
\end{equation}
the absolute amplitude of the wave, and $\omega_{r,n}^{\alpha}, \Gamma_n^{\alpha}$ are the real frequency and growth/damping rate:
\begin{equation}
\omega_{n}^{\alpha} (k) = \omega_{r,n}^{\alpha} (k) + i \Gamma_n^{\alpha} (k),
\end{equation}
with
$\omega_{r,n}^{\alpha} (-k) = - \omega_{r,n}^{\alpha} (k)$
and
$\Gamma_n^{\alpha} (-k) = \Gamma_n^{\alpha} (k)$.
Because the magnetic field is real, 
$\phi_n^{\alpha} (-k) = - \phi_n^{\alpha} (k)$
and
$\delta A_{n}^{\alpha}(-k, x, t) = \delta A_{n}^{\alpha *}(k, x, t)$.

Assuming the waves have random phases,
\begin{equation}
\left\langle A_n^{\alpha} (k) A_m^{\beta *} (k') \right\rangle = 
|A_n^{\alpha}(k)|^2 \delta_{nm} \delta_{\alpha \beta} \delta (k - k'),
\end{equation}
where the brackets here $\langle \cdot \rangle$ indicate average on an ensemble of realizations.
From the 
Maxwell-Faraday equation relating the electromagnetic fields of the transverse waves
\begin{equation}
\frac{\partial \mathbf{B}}{\partial t} = - c \nabla \times \mathbf{E},
\end{equation}
we can express the electric field in terms of the waves $A_n^{\alpha}$:
{\footnotesize
\begin{equation}
\begin{split}
\delta E_{y}(x, t) = &\sum_n \int_{-\infty}^{\infty} dk \delta E_{y,n}(k, x, t) \\
= &- \sum_{n} \int_{-\infty}^{\infty} dk \frac{i}{kc} \frac{\partial}{\partial t} \delta B_{z,n} (k, x, t) \\
= &- \sum_{n} \int_{-\infty}^{\infty} dk \frac{i}{2} \sign (l_n k) \left[ - \frac{\omega_n^L(k)}{kc} \delta A_{n}^L (k, x, t) + \right. \\ 
&+ \left. \frac{\omega_n^R(k)}{kc} \delta A_{n}^R (k, x, t) \right],
\end{split}
\end{equation}
\begin{equation}
\begin{split}
\delta E_{z}(x, t) = &\sum_n \int_{-\infty}^{\infty} dk \delta E_{z,n}(k, x, t) \\
= &\sum_{n} \int_{-\infty}^{\infty} dk \frac{i}{kc} \frac{\partial}{\partial t} \delta B_{y,n} (k, x, t) \\
= &\sum_{n} \int_{-\infty}^{\infty} dk \frac{1}{2} \left[ \frac{\omega_n^L(k)}{kc} \delta A_{n}^L (k, x, t) + \right. \\
&+ \left. \frac{\omega_n^{R}(k)}{kc} \delta A_{n}^R (k, x, t) \right].
\end{split}
\end{equation}
}

Now from the Lorentz force on the particle
\begin{equation}
\dot{\mathbf{p}} = \varepsilon e \left[ \mathbf{\delta E} + \frac{1}{c} \mathbf{v \times \left(B_0 + \delta B \right)} \right],
\end{equation}
it is straightforward to show that
\begin{multline}
\dot{\mu} 
= \varepsilon \frac{\Omega}{B_0} \sqrt{1 - \mu^2} \times \\ 
\times \left[ \left( \delta B_z - \frac{c \mu}{v} \delta E_y \right) \cos \psi
- \left( \delta B_y + \frac{c\mu}{v} \delta E_z \right) \sin \psi\right].
\label{eq:dmudt_full}
\end{multline}

In the environments we are interested in, the phase-speed of the waves $\omega_{n,r}^{\alpha}/k \approx v_A \ll c$. Therefore 
we neglect the contribution of the electric field in~\ref{eq:dmudt_full}:
\begin{multline}
\dot{\mu} = \varepsilon \frac{\Omega}{2 B_0} \sqrt{1 - \mu^2} \times \\
\times \bigl\{ \delta B_z   \left[ \exp (  i \psi) + \exp (- i \psi) \right]
+ \delta B_y i \left[ \exp (  i \psi) - \exp (- i \psi) \right] \bigr\}.
\label{eq:dmudt}
\end{multline}

Using~\ref{eq:dmudt} in combination with \ref{eq:orbit}, \ref{eq:by}, and \ref{eq:bz}:
\begin{multline}
\left\langle \dot{\mu}'(s) \dot{\mu}'(s + \tau) \right\rangle  =
\frac{\Omega^2}{2 B_0^2} \left( 1 - \mu^2 \right) \sum_n \int_{0}^{\infty} dk \times \\
\times \Re \biggl\{ | A_n^{L}(k,s) |^2 \exp \left\{ - i \left[ k v \mu - \omega_n^{L*}(k) + \sign (l_n) \varepsilon \Omega \right] \tau \right\} + \biggr. \\
\biggl. + | A_n^{R}(k,s) |^2 \exp \left\{ - i \left[ k v \mu - \omega_n^{R*}(k) - \sign (l_n) \varepsilon \Omega \right] \tau \right\} \biggr\} . \\
\label{eq:dmudt_dmudt}
\end{multline}

Then integrating~\ref{eq:dmudt_dmudt},
\begin{multline}
\int_{0}^{t} d \tau \left\langle \dot{\mu}'(s) \dot{\mu}'(s + \tau) \right\rangle  =
\frac{\Omega^2}{2 B_0^2} \left( 1 - \mu^2 \right) \sum_n \int_{0}^{\infty} dk \times \\
\times \left\{ | A_n^{L}(k,s) |^2 \mathcal{R} (n, L, k, t)
+ | A_n^{R}(k,s) |^2 \mathcal{R} (n, R, k, t) \right\} \\
\label{eq:int_dmudt_dmudt}
\end{multline}
where the resonance function $\mathcal{R}$ 
(\citealt{schlickeiser_2002, weidl_etal_2015}) 
is defined by:
\begin{multline}
\mathcal{R} (n, \alpha, k, t) \equiv
\Re \biggl\{ \frac{1 - \exp \left[ \Gamma_n^{\alpha}(k) t \right] \exp \left[ - i \vartheta_n^{\alpha} (k) t \right]}{\left[ - \Gamma_n^{\alpha}(k) + i \vartheta_n^{\alpha}(k) \right]} \biggr\} \\
= \frac{- \Gamma_n^{\alpha} \left[ 1 - \exp \left( \Gamma_n^{\alpha} t \right) \cos \left( \vartheta_n^{\alpha} t \right) \right] 
+ \vartheta_n^{\alpha} \exp \left( \Gamma_n^{\alpha} t \right) \sin \left( \vartheta_n^{\alpha} t \right)}{\left( \Gamma_n^{\alpha} \right)^2 + \left( \vartheta_n^{\alpha} \right)^2},
\label{eq:res_func}
\end{multline}
where
\begin{equation}
\vartheta_n^{\alpha} (k) \equiv k v \mu - \omega_{r,n}^{\alpha}(k) - P(\alpha) \sign (l_n) \varepsilon \Omega,
\label{eq:vartheta}
\end{equation}
and $P(\alpha)$ is the polarization ($-1$ for $\alpha=L$ and $+1$ for $\alpha=R$).

For constant amplitude waves ($\Gamma_n^{\alpha} (k) \to 0$),

\begin{equation}
\lim_{\Gamma \to 0} \mathcal{R} (n, \alpha, k, t) = \frac{\sin \left[ \vartheta_n^{\alpha} (k) t \right]}{\vartheta_n^{\alpha}(k)},
\end{equation}
and the main contribution to the integral comes from $|\vartheta_n^{\alpha}(k) t| \le \pi$.
In the limit $t\to\infty$,
\begin{equation}
\lim_{t \to \infty} \left\{ \lim_{\Gamma \to 0} \mathcal{R} (n, \alpha, k, t) \right\} = \pi \delta \left[ \vartheta_n^{\alpha} (k) \right].
\end{equation}

For finite $\Gamma_n^{\alpha} (k) < 0$ (damped waves), 
\begin{equation}
\lim_{t \to \infty}\mathcal{R} (n, \alpha, k, t) = \frac{- \Gamma_n^{\alpha}(k)}{\left[ \Gamma_n^{\alpha}(k) \right]^2 + \left[ \vartheta_n^{\alpha}(k) \right]^2} \;\;\; (\Gamma_n^{\alpha} (k) < 0),
\end{equation}
from which the limit $\Gamma \to 0$ gives again
\begin{equation}
\lim_{\Gamma \to 0^{-}} \left\{ \lim_{t \to \infty} \mathcal{R} (n, \alpha, k, t) \right\} = \pi \delta \left[ \vartheta_n^{\alpha} (k) \right].
\end{equation}

We are interested in evaluating the diffusion during time intervals short compared to the growth/damping time of the wave, so that $| \Gamma_n^{\alpha}(k) t | \ll 1$. 
Keeping only first order terms in $\Gamma_n^{\alpha}(k) t$,

\begin{multline}
\mathcal{R} (n, \alpha, k, t) = \\
= \frac{\left( \Gamma_n^{\alpha} \right)^2 t \cos \left( \vartheta_n^{\alpha} t \right) + \vartheta_n^{\alpha} \left(1 + \Gamma_n^{\alpha} t \right) \sin \left( \vartheta_n^{\alpha} t \right)}{\left( \Gamma_n^{\alpha} \right)^2 + \left( \vartheta_n^{\alpha} \right)^2}
+ \mathcal{O} \left\{ \left( \Gamma_n^{\alpha} t \right)^2 \right\}
\label{eq:r_full}
\end{multline}

For the wavenumbers where $|\vartheta_n^{\alpha}(k) t| > |\Gamma_{n}^{\alpha}(k) t|$,
\begin{multline}
\mathcal{R} (n, \alpha, k, t) 
= \left( 1 + \Gamma_n^{\alpha} t \right) \frac{\sin \left[ \vartheta_n^{\alpha} t \right]}{\vartheta_n^{\alpha}} + \\
+ \mathcal{O} \left\{ \left( \Gamma_n^{\alpha} / \vartheta_n^{\alpha} \right)^2 \right\} + \mathcal{O} \left\{ \left( \Gamma_n^{\alpha} t \right)^2 \right\}.
\label{eq:r_vartheta_gt_gamma}
\end{multline}

Keeping $t$ fixed, 
the above expression shows the main contribution of the resonance function $\mathcal{R}$ in the integral
~\ref{eq:int_dmudt_dmudt} comes from the wavenumbers with $|\vartheta_n^{\alpha} (k) t| \le \pi$. 
Neglecting the term $\omega_{r,n}^{\alpha} (k)$ in the expression for $\vartheta_n^{\alpha} (k)$ (Eq.~\ref{eq:vartheta}),
it gives the following relative window in wave-numbers:
\begin{equation}
\frac{\Delta k}{|k_{\rm res}|} = \frac{2 \pi}{\Omega t} = \gamma \frac{2 \pi}{\Omega_0 t},
\end{equation}
where $\vartheta_n^{\alpha} (k_{\rm res}) = 0$, and $\Omega_0$ is the non-relativistic cyclo-frequency.

\begin{figure}
\centering 
\includegraphics[width=\columnwidth]{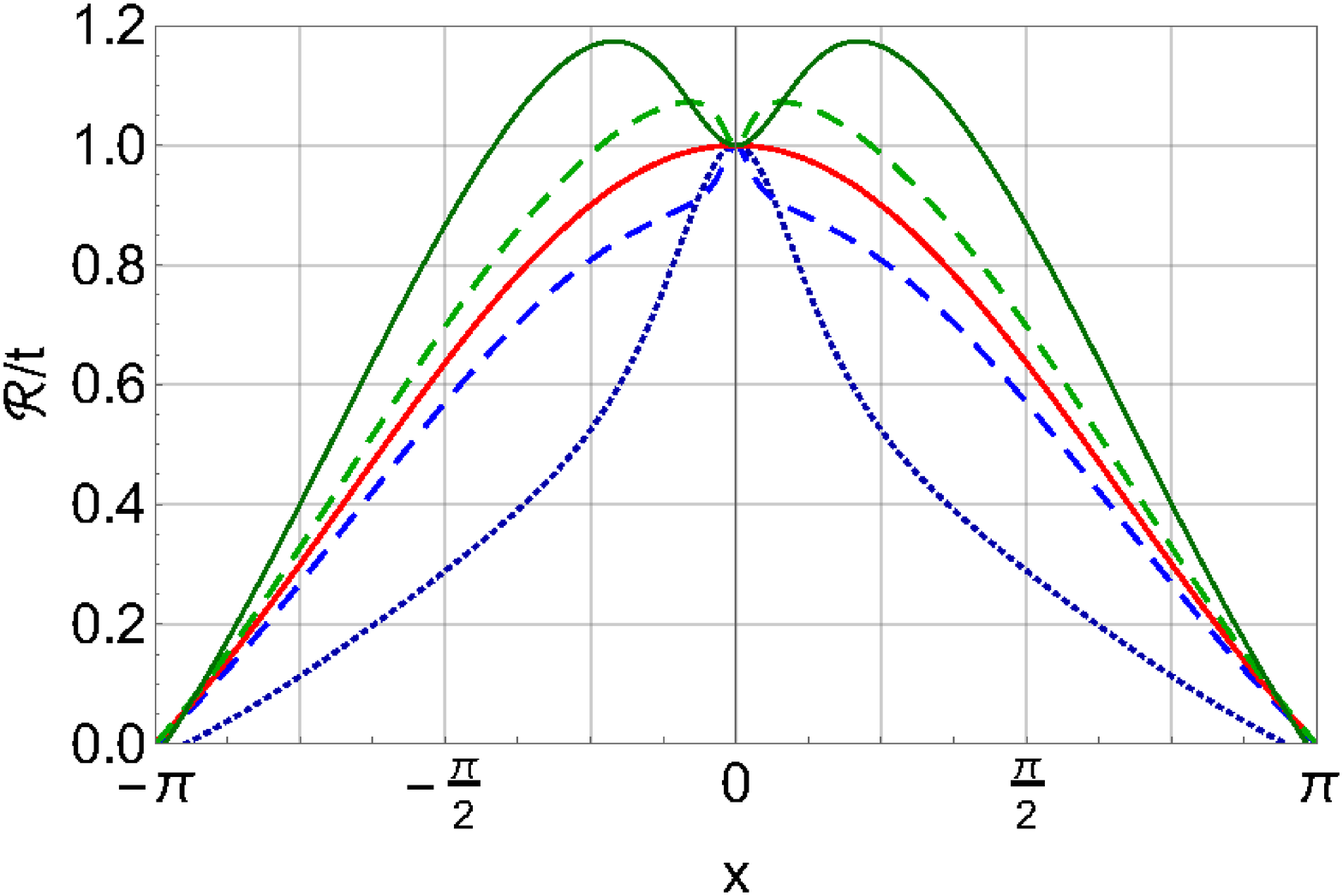}
\caption{The plot of the $\mathcal{R}/t$ defined by Eq.~\ref{eq:R} depending on $x$. 
The values of $c$ is taken to be
(from the lower to the upper curve) 
$-0.5$ ({\color{darkblue} dotted dark blue}), 
$-0.1$ ({\color{blue} dashed blue}), 
$0$ ({\color{red} solid red}), 
$0.1$ ({\color{darkgreen} dashed green}), and 
$0.5$ ({\color{darkgreen2} dotted dark green}).}
\label{fig:Rt}
\end{figure}

Assuming the fixed $t$ is long enough so that $\Delta k / |k_{\rm res}| \ll 1$, 
we can approximate $\Gamma_n^{\alpha} (k) \approx \Gamma_n^{\alpha} (k_{\rm res})$ inside the ressonance window.
Using the notation $c = \Gamma_n^{\alpha}(k_{\rm res}) t$, $x = \vartheta_n^{\alpha} t$, Eq.~\ref{eq:r_full} becomes:
\begin{equation}
\mathcal{R} (n, \alpha, k, t) = t \left\{ \frac{c^2 \cos x + x (1 + c) \sin x}{c^2 + x^2} \right\}.
\label{eq:R}
\end{equation}
Figure~\ref{fig:Rt} shows the behaviour of $\mathcal{R}/t$ in the interval $-\pi \le x \le \pi$ for different 
values of $c$ ($-0.5 \le c \le 0.5$).
Therefore, for growing waves ($\Gamma_n^{\alpha} > 0$), 
\begin{equation}
\mathcal{R} (n, \alpha, k, t) \ge \frac{\sin \left[ \vartheta_n^{\alpha} (k) t \right]}{\vartheta_n^{\alpha}(k)}.
\end{equation}

Now returning to Eq.~\ref{eq:int_dmudt_dmudt} and using the notation
\begin{equation}
\overline{ | A_n^{\alpha} (k,t) |^2 } \equiv \frac{1}{t} \int_{0}^{t} ds |A_n^{\alpha} (k,s)|^2,
\end{equation}
and
\begin{equation}
\widetilde{D}_{\mu\mu} (t)
= \frac{\Omega^2}{2 B_0^2} \left( 1 - \mu^2 \right)
\sum_{n, \alpha} \int_{0}^{\infty} dk \overline{ | A_n^{\alpha} (k,t) |^2 } \frac{\sin \left[ \vartheta_n^{\alpha} (k) t \right]}{\vartheta_n^{\alpha} (k)},
\label{eq:dmumu_min}
\end{equation}
we obtain (for growing waves):
\begin{equation}
D_{\mu\mu} (t) \ge \widetilde{D}_{\mu\mu} (t).
\label{eq:dmumu_tilde_ineq}
\end{equation}
%


\bsp	
\label{lastpage}
\end{document}

%% file: figs/anis_logp_s1-low.tex
\begingroup
  \makeatletter
  \providecommand\color[2][]{%
    \GenericError{(gnuplot) \space\space\space\@spaces}{%
      Package color not loaded in conjunction with
      terminal option `colourtext'%
    }{See the gnuplot documentation for explanation.%
    }{Either use 'blacktext' in gnuplot or load the package
      color.sty in LaTeX.}%
    \renewcommand\color[2][]{}%
  }%
  \providecommand\includegraphics[2][]{%
    \GenericError{(gnuplot) \space\space\space\@spaces}{%
      Package graphicx or graphics not loaded%
    }{See the gnuplot documentation for explanation.%
    }{The gnuplot epslatex terminal needs graphicx.sty or graphics.sty.}%
    \renewcommand\includegraphics[2][]{}%
  }%
  \providecommand\rotatebox[2]{#2}%
  \@ifundefined{ifGPcolor}{%
    \newif\ifGPcolor
    \GPcolortrue
  }{}%
  \@ifundefined{ifGPblacktext}{%
    \newif\ifGPblacktext
    \GPblacktextfalse
  }{}%
  \let\gplgaddtomacro\g@addto@macro
  \gdef\gplbacktext{}%
  \gdef\gplfronttext{}%
  \makeatother
  \ifGPblacktext
    \def\colorrgb#1{}%
    \def\colorgray#1{}%
  \else
    \ifGPcolor
      \def\colorrgb#1{\color[rgb]{#1}}%
      \def\colorgray#1{\color[gray]{#1}}%
      \expandafter\def\csname LTw\endcsname{\color{white}}%
      \expandafter\def\csname LTb\endcsname{\color{black}}%
      \expandafter\def\csname LTa\endcsname{\color{black}}%
      \expandafter\def\csname LT0\endcsname{\color[rgb]{1,0,0}}%
      \expandafter\def\csname LT1\endcsname{\color[rgb]{0,1,0}}%
      \expandafter\def\csname LT2\endcsname{\color[rgb]{0,0,1}}%
      \expandafter\def\csname LT3\endcsname{\color[rgb]{1,0,1}}%
      \expandafter\def\csname LT4\endcsname{\color[rgb]{0,1,1}}%
      \expandafter\def\csname LT5\endcsname{\color[rgb]{1,1,0}}%
      \expandafter\def\csname LT6\endcsname{\color[rgb]{0,0,0}}%
      \expandafter\def\csname LT7\endcsname{\color[rgb]{1,0.3,0}}%
      \expandafter\def\csname LT8\endcsname{\color[rgb]{0.5,0.5,0.5}}%
    \else
      \def\colorrgb#1{\color{black}}%
      \def\colorgray#1{\color[gray]{#1}}%
      \expandafter\def\csname LTw\endcsname{\color{white}}%
      \expandafter\def\csname LTb\endcsname{\color{black}}%
      \expandafter\def\csname LTa\endcsname{\color{black}}%
      \expandafter\def\csname LT0\endcsname{\color{black}}%
      \expandafter\def\csname LT1\endcsname{\color{black}}%
      \expandafter\def\csname LT2\endcsname{\color{black}}%
      \expandafter\def\csname LT3\endcsname{\color{black}}%
      \expandafter\def\csname LT4\endcsname{\color{black}}%
      \expandafter\def\csname LT5\endcsname{\color{black}}%
      \expandafter\def\csname LT6\endcsname{\color{black}}%
      \expandafter\def\csname LT7\endcsname{\color{black}}%
      \expandafter\def\csname LT8\endcsname{\color{black}}%
    \fi
  \fi
  \setlength{\unitlength}{0.0500bp}%
  \begin{picture}(4896.00,3168.00)%
    \gplgaddtomacro\gplbacktext{%
      \csname LTb\endcsname%
      \put(857,660){\makebox(0,0)[r]{\strut{}$-0.2$}}%
      \put(857,1177){\makebox(0,0)[r]{\strut{}$-0.1$}}%
      \put(857,1694){\makebox(0,0)[r]{\strut{}$0.0$}}%
      \put(857,2210){\makebox(0,0)[r]{\strut{}$0.1$}}%
      \put(857,2727){\makebox(0,0)[r]{\strut{}$0.2$}}%
      \put(989,440){\makebox(0,0){\strut{}$-0.5$}}%
      \put(1850,440){\makebox(0,0){\strut{}$0.0$}}%
      \put(2712,440){\makebox(0,0){\strut{}$0.5$}}%
      \put(3573,440){\makebox(0,0){\strut{}$1.0$}}%
      \put(4434,440){\makebox(0,0){\strut{}$1.5$}}%
      \put(219,1693){\rotatebox{-270}{\makebox(0,0){\strut{}$P_{\perp}/P_{\parallel} - 1$}}}%
      \put(2711,220){\makebox(0,0){\strut{}$\log \left( p / p_{\min} \right)$}}%
      \put(2711,3057){\makebox(0,0){$A_0 = +0.1$}}%
    }%
    \gplgaddtomacro\gplfronttext{%
    }%
    \gplbacktext
    \put(0,0){\includegraphics{anis_logp_s1-low}}%
    \gplfronttext
  \end{picture}%
\endgroup

%% file: figs/anis_logp_s5-low.tex
\begingroup
  \makeatletter
  \providecommand\color[2][]{%
    \GenericError{(gnuplot) \space\space\space\@spaces}{%
      Package color not loaded in conjunction with
      terminal option `colourtext'%
    }{See the gnuplot documentation for explanation.%
    }{Either use 'blacktext' in gnuplot or load the package
      color.sty in LaTeX.}%
    \renewcommand\color[2][]{}%
  }%
  \providecommand\includegraphics[2][]{%
    \GenericError{(gnuplot) \space\space\space\@spaces}{%
      Package graphicx or graphics not loaded%
    }{See the gnuplot documentation for explanation.%
    }{The gnuplot epslatex terminal needs graphicx.sty or graphics.sty.}%
    \renewcommand\includegraphics[2][]{}%
  }%
  \providecommand\rotatebox[2]{#2}%
  \@ifundefined{ifGPcolor}{%
    \newif\ifGPcolor
    \GPcolortrue
  }{}%
  \@ifundefined{ifGPblacktext}{%
    \newif\ifGPblacktext
    \GPblacktextfalse
  }{}%
  \let\gplgaddtomacro\g@addto@macro
  \gdef\gplbacktext{}%
  \gdef\gplfronttext{}%
  \makeatother
  \ifGPblacktext
    \def\colorrgb#1{}%
    \def\colorgray#1{}%
  \else
    \ifGPcolor
      \def\colorrgb#1{\color[rgb]{#1}}%
      \def\colorgray#1{\color[gray]{#1}}%
      \expandafter\def\csname LTw\endcsname{\color{white}}%
      \expandafter\def\csname LTb\endcsname{\color{black}}%
      \expandafter\def\csname LTa\endcsname{\color{black}}%
      \expandafter\def\csname LT0\endcsname{\color[rgb]{1,0,0}}%
      \expandafter\def\csname LT1\endcsname{\color[rgb]{0,1,0}}%
      \expandafter\def\csname LT2\endcsname{\color[rgb]{0,0,1}}%
      \expandafter\def\csname LT3\endcsname{\color[rgb]{1,0,1}}%
      \expandafter\def\csname LT4\endcsname{\color[rgb]{0,1,1}}%
      \expandafter\def\csname LT5\endcsname{\color[rgb]{1,1,0}}%
      \expandafter\def\csname LT6\endcsname{\color[rgb]{0,0,0}}%
      \expandafter\def\csname LT7\endcsname{\color[rgb]{1,0.3,0}}%
      \expandafter\def\csname LT8\endcsname{\color[rgb]{0.5,0.5,0.5}}%
    \else
      \def\colorrgb#1{\color{black}}%
      \def\colorgray#1{\color[gray]{#1}}%
      \expandafter\def\csname LTw\endcsname{\color{white}}%
      \expandafter\def\csname LTb\endcsname{\color{black}}%
      \expandafter\def\csname LTa\endcsname{\color{black}}%
      \expandafter\def\csname LT0\endcsname{\color{black}}%
      \expandafter\def\csname LT1\endcsname{\color{black}}%
      \expandafter\def\csname LT2\endcsname{\color{black}}%
      \expandafter\def\csname LT3\endcsname{\color{black}}%
      \expandafter\def\csname LT4\endcsname{\color{black}}%
      \expandafter\def\csname LT5\endcsname{\color{black}}%
      \expandafter\def\csname LT6\endcsname{\color{black}}%
      \expandafter\def\csname LT7\endcsname{\color{black}}%
      \expandafter\def\csname LT8\endcsname{\color{black}}%
    \fi
  \fi
  \setlength{\unitlength}{0.0500bp}%
  \begin{picture}(4896.00,3168.00)%
    \gplgaddtomacro\gplbacktext{%
      \csname LTb\endcsname%
      \put(857,660){\makebox(0,0)[r]{\strut{}$-0.4$}}%
      \put(857,1177){\makebox(0,0)[r]{\strut{}$-0.2$}}%
      \put(857,1694){\makebox(0,0)[r]{\strut{}$0.0$}}%
      \put(857,2210){\makebox(0,0)[r]{\strut{}$0.2$}}%
      \put(857,2727){\makebox(0,0)[r]{\strut{}$0.4$}}%
      \put(989,440){\makebox(0,0){\strut{}$-0.5$}}%
      \put(1850,440){\makebox(0,0){\strut{}$0.0$}}%
      \put(2712,440){\makebox(0,0){\strut{}$0.5$}}%
      \put(3573,440){\makebox(0,0){\strut{}$1.0$}}%
      \put(4434,440){\makebox(0,0){\strut{}$1.5$}}%
      \put(219,1693){\rotatebox{-270}{\makebox(0,0){\strut{}$P_{\perp}/P_{\parallel} - 1$}}}%
      \put(2711,220){\makebox(0,0){\strut{}$\log \left( p / p_{\min} \right)$}}%
      \put(2711,3057){\makebox(0,0){$A_0 = -0.3$}}%
    }%
    \gplgaddtomacro\gplfronttext{%
    }%
    \gplbacktext
    \put(0,0){\includegraphics{anis_logp_s5-low}}%
    \gplfronttext
  \end{picture}%
\endgroup